\documentclass[aps,prl,amsmath,amssymb, preprintnumbers, reprint,longbibliography,superscriptaddress]{revtex4-1}
\pdfoutput=1
\usepackage{lmodern, graphicx, multirow, xcolor, adjustbox}
\usepackage[colorlinks=true, citecolor=blue, urlcolor=blue, linkcolor=blue, breaklinks=true, pdfpagelabels=false]{hyperref}

\makeatletter\g@addto@macro\bfseries{\boldmath}\makeatother
\DeclareMathSymbol{\shortminus}{\mathbin}{AMSa}{"39}

\catcode`\$=\active
\gdef$#1${\texorpdfstring{\(#1\)}{\detokenize{#1}}}

\renewcommand{\phi}{\ensuremath{\varphi}}
\DeclareMathOperator{\Tr}{Tr}
\allowdisplaybreaks[4]
\newcommand{\OO}{O}
\newcommand{\cc}{c}
\newcommand{\obs}{o}
\newcommand{\sss}{\scriptscriptstyle}
\newcommand{\err}[4]{#1(#2)^{\sss#3}_{\sss#4}}
\newcommand{\erro}[4]{#1^{\sss#3}_{\sss#4}}

\begin{document}

\newcommand{\cpthree}{Centre for Cosmology, Particle Physics and Phenomenology (CP3),  Universit\'{e} catholique de Louvain, 1348 Louvain-la-Neuve, Belgium}

\author{C\'eline Degrande}
\email{celine.degrande@uclouvain.be}
\affiliation{\cpthree}

\author{Gauthier Durieux}
\email{gauthier.durieux@cern.ch}
\affiliation{Physics Department, Technion -- Israel Institute of Technology, Haifa 3200003, Israel}

\author{Fabio Maltoni}
\email{fabio.maltoni@uclouvain.be, fabio.maltoni@unibo.it}
\affiliation{\cpthree}
\affiliation{Dipartimento di Fisica e Astronomia, Universit\`a di Bologna e INFN, Sezione di Bologna, via Irnerio 46, I-40126 Bologna, Italy}

\author{Ken Mimasu}
\email{ken.mimasu@uclouvain.be}
\affiliation{\cpthree}

\author{Eleni Vryonidou}
\email{eleni.vryonidou@cern.ch}
\affiliation{CERN, Theoretical Physics Department, Geneva 23 CH-1211, Switzerland}

\author{Cen Zhang}
\email{cenzhang@ihep.ac.cn}
\affiliation{Institute for High Energy Physics, and School of Physical Sciences, University
of Chinese Academy of Sciences, Beijing 100049, China}
\affiliation{Center for High Energy Physics, Peking University, Beijing 100871, China}

\preprint{CERN-TH-2020-140, CP3-20-42}

\title{Automated one-loop computations in the SMEFT}

\begin{abstract}
We present the automation of one-loop computations in the standard-model effective field theory at dimension six.
Our general implementation, dubbed {\sc SMEFT@NLO}, covers all types of operators: bosonic, two- and four-fermion ones.
Included ultraviolet and rational counterterms presently allow for fully differential predictions, possibly matched to parton shower, up to the one-loop level in the strong coupling or in four-quark operator coefficients.
Exact flavor symmetries are imposed among light quark generations and an initial focus is set on top-quark interactions in the fermionic sector.
We illustrate the potential of this implementation with novel loop-induced and next-to-leading-order computations relevant for top-quark, electroweak, and Higgs-boson phenomenology at the LHC and future colliders.
\end{abstract}

\maketitle

\subparagraph{Introduction}
Observed deviations in accurate measurements would indirectly point to the existence of physics beyond the standard model (SM), even if heavy new states remain out of reach of the LHC and foreseen accelerators.
Given the richness of collider observables and of the models proposed to address SM limitations, a clear strategy is needed to maximize the reach of present and future experiments.

The standard-model effective field theory (SMEFT) provides a powerful framework to {\it search for} and {\it interpret} possible deviations from the SM~\cite{Weinberg:1978kz,Buchmuller:1985jz,Leung:1984ni}.
Its use is complementary to direct searches.
Higher-dimensional operators compatible with the symmetries of the SM generate a well-defined pattern of new interaction terms.
Their relevance is dictated, \emph{a priori}, by the operator dimension, \emph{i.e.}, by an expansion in $1/\Lambda$,
\begin{equation}
{\cal L}_{\textrm{SMEFT}} = 
{\cal L}_{\textrm{SM}} + {\sum}_i
\frac{c^{(6)}_i  \OO_i^{(6)}}{\Lambda^2} + {\cal O}\left(\frac{1}{\Lambda^{3}}\right)\,,
\end{equation}
where ${\cal L}_{\textrm{SM}}$ is the SM Lagrangian, $\OO_i^{(d)}$ are operators of dimension $d$ larger than four, and the $c_i^{(d)}$ are the corresponding Wilson coefficients which encode  information about the ultraviolet (UV) theory.
We do not consider the single operator of dimension five which violates lepton number and generates Majorana neutrino masses.
At dimension six, without considering the combinatorial complexity introduced by non-trivial flavor structures, the number of independent operators is remarkably small~\cite{Henning:2015alf}.
Just 84 parameters encode the leading indirect effects from \emph{all} flavor-blind scenarios of decoupling new physics.

One can then parametrize possible deviations from the SM prediction, for \emph{any} observable $\obs_n$, in terms of the Wilson coefficients
\begin{equation}
    \Delta \obs_n
    =
    \obs_n^{\text{EXP}}\!\! - \obs_n^{\text{SM}} 
     =
   {\sum}_i
      \dfrac{a^{\scriptstyle(6)}_{n,i}(\mu) \,  c^{\scriptstyle(6)}_i(\mu)}{\Lambda^2} 
    + {\cal O}\left(\frac{1}{\Lambda^{3}}\right) \,,
\end{equation}
where $\obs_n^{\textrm{SM}}$ and $a^{(6)}_{n,i}$ are calculated using standard techniques as expansions in the strong and weak couplings, while $\mu$ is the renormalization scale.
The expression above illustrates the key points of a precision approach to the search for new physics.
First, one needs to achieve the highest precision in both the experimental and SM determinations of the observables $\obs_n$ to reliably identify the corresponding deviation $\Delta\obs_n$.
Second, since the SMEFT correlates these deviations, improving its predictions enhances our sensitivity to new-physics patterns.
Third, in presence of a signal, the identification of the UV physics based on the extracted $c^{(6)}_i/\Lambda^2$ can be greatly affected by the accuracy and precision on the $a^{(6)}_{n,i}$.
Hence, to fully exploit the measurements, it is not only mandatory to have the best SM calculations but also to control the accuracy and uncertainties of the SMEFT predictions.
In this article, we present an important milestone in this direction, allowing to automatically compute higher-order contributions to SMEFT predictions, for any observable of interest.

\subparagraph{Generalities}
Adopting the \emph{Warsaw} basis~\cite{Grzadkowski:2010es} and after canonical normalization, we implement dimension-six SMEFT operators in a {\sc FeynRules}~\cite{Alloul:2013bka} model dubbed {\sc SMEFT@NLO}. 
This implementation is publicly available online together with its technical documentation, including operator definitions~\cite{smeft:2020}.

We employ $G_F$, $m_Z$ and $m_W$ as electroweak input parameters so that propagators do not depend on operator coefficients.
A linear expansion of Feynman rules is therefore sufficient to perform an exact truncation of matrix elements to leading SMEFT order in Monte-Carlo programs.
Given the invariance of the $S$-matrix under field redefinitions, results at that order can be translated exactly from one dimension-six operator basis to another.

We consider flavor structures relevant for collider observables and new physics that might single out the top quark.
The Cabibbo-Kobayashi-Maskawa matrix is approximated as a unit matrix.
All fermion masses and Yukawa couplings are neglected, except that of the top quark.
An $U(2)_q\times U(2)_u\times U(3)_d$ flavor symmetry is imposed among the first two generations of left-handed quark doublets and up-type right-handed singlets as well as among all three generations of right-handed down-type singlets.
Chirality flipping and right-handed charged currents involving light-quarks ---right-handed bottom included--- are thus forbidden.
This guarantees consistency with the five-flavor scheme we adopt, where the bottom quark is approximated as massless to avoid the generation of large logarithms of the ratio of $m_b$ to the hard scale of the process considered.
In the current implementation, we moreover focus primarily on operators involving a top quark.
In the lepton sector, we enforce a $[U(1)_l \times U(1)_e]^3$ symmetry which results in flavor diagonality and forbids chirality flipping interactions.
Where relevant, our notation and normalizations match those of Ref.~\cite{AguilarSaavedra:2018nen}.

Once passed to {\sc MadGraph5\_aMC@NLO}~\cite{Alwall:2014hca}, the model allows the \emph{tree-level} calculation of observables at any order in SM couplings, with the possibility of evaluating the contributions that are linear and quadratic in the Wilson coefficients separately or that involve multiple operator insertions.
After linear transformations between conventions, tree-level results for amplitudes computed at individual phase-space points match that of other implementations~\cite{Brivio:2017btx, AguilarSaavedra:2018nen} to machine precision~\cite{Durieux:2019lnv}.

\subparagraph{One-loop computations}
Automating one-loop calculations requires both UV and rational counterterms. The former encode the renormalisation of Lagrangian parameters
while the latter are required to palliate the numerical treatment of the Dirac algebra in four dimensions~\cite{Ossola:2006us}.
Depending on the complexity of the theory, their determination can become tedious and cumbersome.
Yet, being process independent, they only need to be computed once and for all.

Masses and wave-functions are renormalized on-shell while the strong coupling and operator coefficients are treated in the $\overline{\text{MS}}$ scheme.
The generator takes coefficients as input  (possibly renormalization-group-evolved~\cite{Jenkins:2013zja,Jenkins:2013wua,Alonso:2013hga}) and keeps them fixed at a scale distinguished from that of the strong coupling.
The computation of the counterterms necessary for QCD and four-quark operator loops has been performed with an in-house version of the {\sc NLOCT} package~\cite{Degrande:2014vpa} which has been extended to handle the diversity of structures (\textit{e.g.}, Lorentz and color) as well as the higher rank integrands appearing in the SMEFT.

Two particular difficulties arise, associated to the loop-level generation of gauge anomalies by SMEFT modifications of chiral interactions, and to evanescent operators that vanish in four dimensions.
To preserve the QCD Ward identity, the \emph{covariant anomaly} scheme~\cite{Bardeen:1984pm, Fox:2018ldq} has been adopted such that rational counterterms cancel the anomalies in three- and four-point amplitudes such as $ggZ$, $gggZ$, $ggZH$ and $gg\bar\psi\psi$.
In the latter case, the anomaly is generated by four-fermion operators with an axial quark-current closed in a loop to which two gluons are attached.
Since the SMEFT covers heavy new-physics scenarios in which the full SM gauge symmetry is preserved, a matching computation in the same scheme would always result in the anomaly cancellation we require.
In $d$ dimensions, $\gamma_5$ is taken as anticommuting and the cyclic property of traces of Dirac matrices is abandoned~\cite{Kreimer:1989ke, Korner:1991sx, Kreimer:1993bh}.

Evanescent operators~\cite{Dugan:1990df, Herrlich:1994kh} arise in one-loop computations involving four-fermion operators in $D=4-2\epsilon$ dimensions and are required for our implementation of QCD corrections to four-quark operators.
The Dirac algebra is only closed in four dimensions, and a basis of four-quark operators in $D$ dimensions contains an infinite number of operators.
These can for instance be written with antisymmetric products of Dirac matrices~\cite{Dugan:1990df}: $\bar\psi_1\gamma^{[\mu_1}\gamma^{\mu_2}\cdots\gamma^{\mu_n]}\psi_2  \:\;\bar\psi_3\gamma_{[\mu_1}\gamma_{\mu_2}\cdots\gamma_{\mu_n]}\psi_4$.
All such operators with $n\ge5$ are proportional to $\epsilon$ and vanish in four dimensions.
They can however give rise to finite contributions at the one-loop level when they are generated with a coefficient diverging like $1/\epsilon$.
A basis of evanescent operators must therefore be defined.
Although one-loop SMEFT results depend on this choice, employing the same evanescent operator basis in the one-loop matching of the SMEFT to a specific UV model would lead to the cancellation of this arbitrary dependence.
It can therefore be conceived as a scheme which affects our rational counterterms.

We follow the evanescent operator conventions of Ref.~\cite{Buras:1989xd}, used in existing one-loop QCD computations to LHC processes involving four-quark operators~\cite{Gao:2011ha, Shao:2011wa}.
Namely, the $\epsilon$ part of the decomposition of each four-fermion Dirac structure is fixed by matching the trace of the latter with that of its decomposition, when contracted with any basis element.
Separating the Dirac structures of the two fermion bilinears by a $\otimes$ sign, one for example defines the evanescent operator $E$ through
\begin{multline}
\gamma^\mu\gamma^\nu\gamma^\rho\gamma_+ \otimes \gamma_\mu\gamma_\nu\gamma_\rho\gamma_+ 
= E + \sum_k(f_k+a_k\epsilon) \Gamma^k \otimes {\Gamma'}^k\\
= E + (16-4 \epsilon)\gamma^\alpha\gamma_+ \otimes \gamma_\alpha\gamma_+
\end{multline}
where the coefficients of the decomposition, $f_k$ and $a_k$, are obtained by requesting that
\begin{multline}
\Tr\left(\gamma^\mu\gamma^\nu\gamma^\rho\gamma_+ \Gamma^m \gamma_\mu\gamma_\nu\gamma_\rho\gamma_+  {\Gamma'}^m \right) \\= \sum_k(f_k+a_k\epsilon)
\Tr\left(\Gamma^k \Gamma^m  {\Gamma'}^k {\Gamma'}^m\right)
	+\mathcal{O}(\epsilon^2)
\end{multline}
for each element $\Gamma^m \otimes {\Gamma'}^m$ of the chosen basis of four-fermion Dirac structures in four dimensions.
Our basis is 
\begin{multline}  
\big\{
	\gamma_\pm \otimes \gamma_\pm,\;\:
	\gamma_\pm \otimes \gamma_\mp,\;\:
	\gamma^\alpha\gamma_\pm \otimes \gamma_\alpha\gamma_\pm,\;\:\\
	\gamma^\alpha\gamma_\pm \otimes \gamma_\alpha\gamma_\mp,\;\:
	\sigma^{\alpha\beta}\gamma_\pm \otimes \sigma^{\alpha\beta}\gamma_\pm
\big\}
\end{multline}
with $\gamma_\pm\equiv (1\pm\gamma_5)/2$ and $\sigma^{\alpha\beta}\equiv \frac{i}{2}[\gamma^\alpha,\gamma^\beta]$.

Given our assumptions, no flavor-changing interactions are generated at one loop and the bottom quark remains massless.
The closure of the renormalisation procedure at the level of dimension-six operators is therefore guaranteed if loops with at most one operator insertion are allowed.
While the framework can handle any kind of one-loop amplitudes in the SMEFT, the current version only includes the counterterms (up to five points) required for one-loop computations involving the strong coupling or four-quark operators.
By construction, the infrared structure of the SMEFT is identical to that of the SM.
No additional ingredient is thus required to ensure the cancellation of divergences between real and virtual diagrams in that regime, or to match matrix elements to parton showers.

Being fully automatic, our implementation avoids error-prone manual manipulations.
We validated, against analytical results, various one-loop computations relevant for top-quark processes as well as rational counterterms such as the four-quark ones or those that ensure the cancellation of anomalies.
Gauge invariance and pole cancellation have been tested numerically in a wide range of processes, using available built-in routines.
A list of these processes together with guidelines for their generation is available online~\cite{smeft:2020}.
The 3.1 series of {\sc MadGraph5\_aMC@NLO} is required in order to exploit all implementation features and, \emph{e.g.}, to separately compute the linear and quadratic SMEFT contributions at NLO and to make loop-level predictions for four-fermion operators.

\subparagraph{Applications}
While including entirely new elements, the present implementation is built on earlier NLO developments tailored to specific applications:
top-quark FCNC~\cite{Degrande:2014tta, Durieux:2014xla, TopFCNC},
SMEFT effects in $t \bar t$~\cite{Franzosi:2015osa}, $t\bar t H$ and
$gg \to Hj,HH$~\cite{Maltoni:2016yxb},
$t\bar t Z$  and $gg \to ZH$~\cite{Bylund:2016phk},
$tj$~\cite{Zhang:2016omx,deBeurs:2018pvs},
$gg\to H$ in conjunction with analytic two-loop computations~\cite{Deutschmann:2017qum},
multi-jet~\cite{Hirschi:2018etq}, EW Higgs production~\cite{Degrande:2016dqg},
and $tHj,tZj$~\cite{Degrande:2018fog}. 
Global fits in the top-quark sector~\cite{Brivio:2019ius, Hartland:2019bjb} have recently made use of NLO predictions obtained with a development version of {\sc SMEFT@NLO}.

The number of possible applications is too vast to be presented in a comprehensive way in this article.
We therefore provide selected novel examples relevant for top-quark, electroweak, and Higgs-boson phenomenology at the LHC and future colliders, focusing on the importance of NLO effects.
Numerical results assume $c_i/\Lambda^2=1\,\text{TeV}^{-2}$.
For concision, at $\mathcal{O}(\Lambda^{-4})$, we only quote the $c_ic_j$ dependencies for $i=j$.
Unless otherwise specified, we fix the factorization and both renormalization scales to a common value: the sum of final-state masses divided by two.
Uncertainty envelopes are obtained from the separate variations of renormalization and factorization scales by factors of two up and down and are quoted in percent.
The operator coefficients are not evolved.
Monte Carlo errors on the last significant digit are indicated between parentheses, if they exceed 5\%.
The NLO sets of NNPDF3.0~\cite{Ball:2014uwa} are used as parton distribution functions, with $\alpha_S(M_Z)=0.118$.
LO sets are however employed for tree-level and loop-induced processes.
Other relevant parameters are $m_t=173\,{\rm GeV}$, $m_h=125\,{\rm GeV}$, $m_Z=91.1876\,{\rm GeV}$, $m_W=80.41 \,{\rm GeV}$ and $G_F=1.16637 \times 10^{-5}\,\text{GeV}^{-2}$.

\begin{figure}[tb]
\centering
\includegraphics[width=\columnwidth]{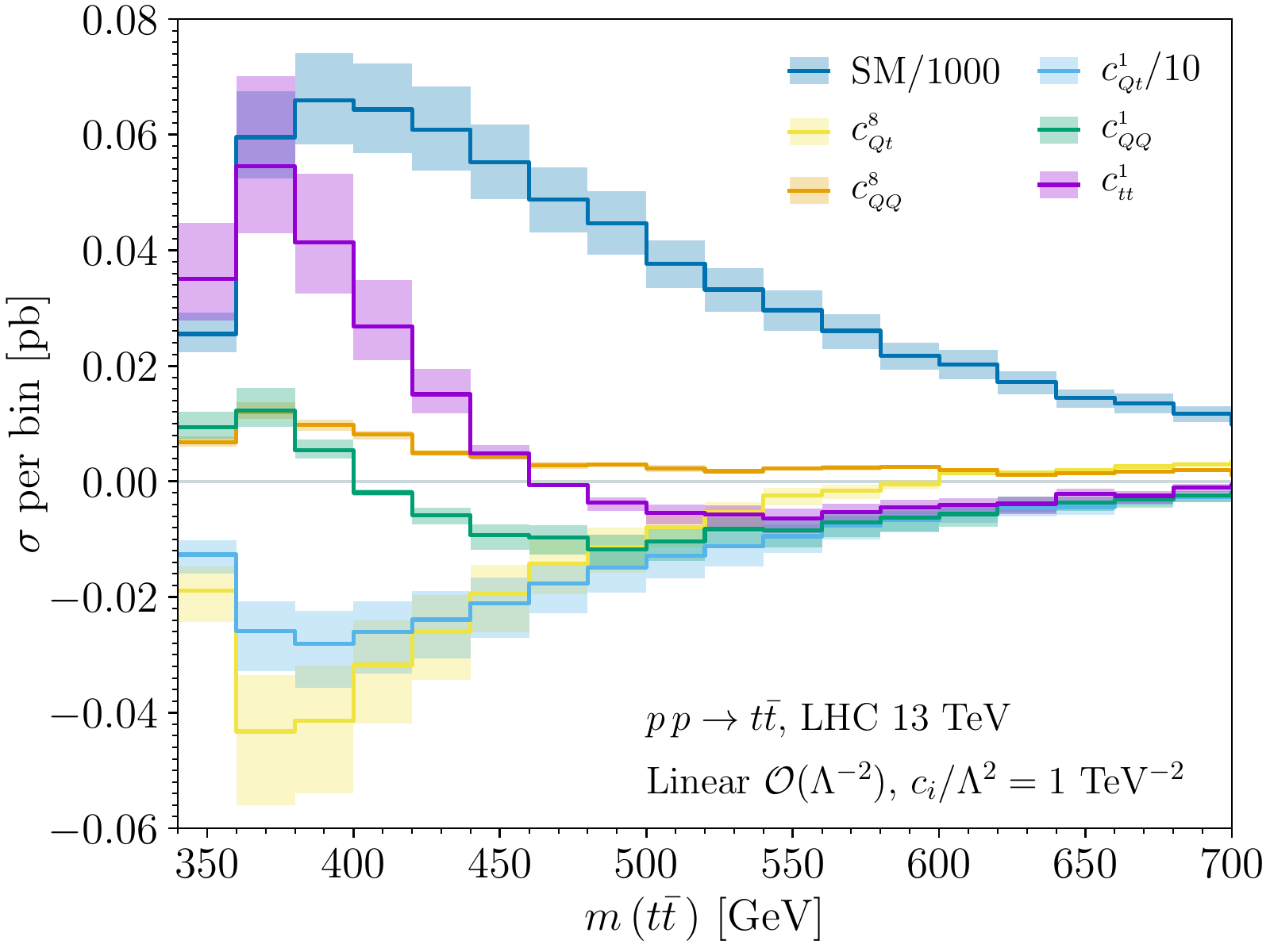}
\caption{$t\bar{t}$ invariant-mass distribution of the interference between four-heavy-quark operators and the SM.
}
\label{fig:mtt}
\end{figure}

As a first application, we present four-fermion operator contributions to top-pair and four-top production. 
Pinning down this sector will provide sensitivity to the well-motivated scenario that new physics couples preferentially to the third generation.
\autoref{tab:ttbar} collects the four-fermion contributions to $t\bar{t}$ production at the LHC $\sqrt{s}=13\,$TeV.
The NLO computation allows us to extract, for the first time, the interference of color-singlet operators with leading QCD contributions.
For $\mathcal{O}(1)$ coefficients, these are typically small, compared to the $\mathcal{O}(\Lambda^{-4})$ terms and to the SM cross-section.
We also compute their interferences up to NLO in QCD with SM electroweak production, which are comparable.
One aspect worth noting is that NLO corrections break the LO degeneracy between various color-octet operators, which could be crucial in global fits, see for instance Ref.~\cite{Brivio:2019ius}.
\begin{table}
\renewcommand{\arraystretch}{1.5}
\newcommand{\mysep}{@{\hspace{1.4mm}}}
\centering
\adjustbox{max width=\columnwidth}{\scriptsize
\begin{tabular}{
    |c
    |c
    |r
    |c
    |c
    |c
    |}
\hline
\multirow{2}{*}{$c_i$}
    &\multicolumn{3}{c\mysep|}{$\mathcal{O}(\Lambda^{-2})$}
    &\multicolumn{2}{c\mysep|}{$\mathcal{O}(\Lambda^{-4})$}    \\[-1mm]
    &\multicolumn{1}{c\mysep }{LO} & \multicolumn{2}{c\mysep|}{NLO}
    &\multicolumn{1}{c\mysep }{LO} & NLO    \\\hline\hline
$\cc_{tu}^{8}$ & $\erro{4.27}{}{+11\%}{-9\%}$ & \multicolumn{2}{c\mysep|}{$\erro{4.06}{}{+1\%}{-3\%}$} & $\erro{1.04}{}{+6\%}{-5\%}$ & $\erro{1.03}{}{+2\%}{-2\%}$\tabularnewline
$\cc_{td}^{8}$ & $\erro{2.79}{}{+11\%}{-9\%}$ & \multicolumn{2}{c\mysep|}{$\erro{2.77}{}{+1\%}{-3\%}$} & $\erro{0.577}{}{+6\%}{-5\%}$ & $\erro{0.611}{}{+3\%}{-2\%}$\tabularnewline
$\cc_{tq}^{8}$ & $\erro{6.99}{}{+11\%}{-9\%}$ & \multicolumn{2}{c\mysep|}{$\erro{6.67}{}{+1\%}{-3\%}$} & $\erro{1.61}{}{+6\%}{-5\%}$ & $\erro{1.29}{}{+3\%}{-2\%}$\tabularnewline
$\cc_{Qu}^{8}$ & $\erro{4.26}{}{+11\%}{-9\%}$ & \multicolumn{2}{c\mysep|}{$\erro{3.93}{}{+1\%}{-4\%}$} & $\erro{1.04}{}{+6\%}{-5\%}$ & $\erro{0.798}{}{+3\%}{-3\%}$\tabularnewline
$\cc_{Qd}^{8}$ & $\erro{2.79}{}{+11\%}{-9\%}$ & \multicolumn{2}{c\mysep|}{$\erro{2.93}{}{+0\%}{-1\%}$} & $\erro{0.58}{}{+6\%}{-5\%}$ & $\erro{0.485}{}{+2\%}{-2\%}$\tabularnewline
$\cc_{Qq}^{8,1}$ & $\erro{6.99}{}{+11\%}{-9\%}$ & \multicolumn{2}{c\mysep|}{$\erro{6.82}{}{+1\%}{-3\%}$} & $\erro{1.61}{}{+6\%}{-5\%}$ & $\erro{1.69}{}{+3\%}{-3\%}$\tabularnewline
$\cc_{Qq}^{8,3}$ & $\erro{1.50}{}{+10\%}{-9\%}$ & \multicolumn{2}{c\mysep|}{$\erro{1.32}{}{+1\%}{-3\%}$} & $\erro{1.61}{}{+6\%}{-5\%}$ & $\erro{1.57}{}{+2\%}{-2\%}$\tabularnewline[0.5mm]
\hline\hline
$\cc_{tu}^{1}$ & $[0.67_{-1\%}^{+1\%}]$ & $\err{\shortminus0.078}{7}{+31\%}{-23\%}$ & $[0.41_{-17\%}^{+13\%}]$ & $\erro{4.66}{}{+6\%}{-5\%}$ & $\erro{5.92}{}{+6\%}{-5\%}$\tabularnewline
$\cc_{td}^{1}$ & $[\shortminus0.21^{+1\%}_{-2\%}]$ & $\erro{\shortminus0.306}{}{+30\%}{-22\%}$ & $[\shortminus0.15^{+10\%}_{-13\%}]$ & $\erro{2.62}{}{+6\%}{-5\%}$ & $\erro{3.46}{}{+5\%}{-5\%}$\tabularnewline
$\cc_{tq}^{1}$ & $[0.39_{-1\%}^{+0\%}]$ & $\erro{\shortminus0.47}{}{+24\%}{-18\%}$ & $[0.50_{-2\%}^{+3\%}]$ & $\erro{7.25}{}{+6\%}{-5\%}$ & $\erro{9.36}{}{+6\%}{-5\%}$\tabularnewline
$\cc_{Qu}^{1}$ & $[0.33_{-0\%}^{+0\%}]$ & $\erro{\shortminus0.359}{}{+23\%}{-17\%}$ & $[0.57^{+6\%}_{-5\%}]$ & $\erro{4.68}{}{+6\%}{-5\%}$ & $\erro{5.96}{}{+6\%}{-5\%}$\tabularnewline
$\cc_{Qd}^{1}$ & $[\shortminus0.11^{+0\%}_{-1\%}]$ & $\err{0.023}{6}{+114\%}{-75\%}$ & $[\shortminus0.19^{+6\%}_{-5\%}]$ & $\erro{2.61}{}{+6\%}{-5\%}$ & $\erro{3.46}{}{+5\%}{-5\%}$\tabularnewline
$\cc_{Qq}^{1,1}$ & $[0.57_{-1\%}^{+0\%}]$ & $\erro{\shortminus0.24}{}{+30\%}{-22\%}$ & $[0.39_{-12\%}^{+9\%}]$ & $\erro{7.25}{}{+6\%}{-5\%}$ & $\erro{9.34}{}{+5\%}{-5\%}$\tabularnewline
$\cc_{Qq}^{1,3}$ & $[1.92_{-1\%}^{+1\%}]$ & $\err{0.088}{7}{+28\%}{-20\%}$ & $[1.05_{-22\%}^{+17\%}]$ & $\erro{7.25}{}{+6\%}{-5\%}$ & $\erro{9.32}{}{+5\%}{-5\%}$\tabularnewline[0.5mm]
\hline\hline
$\cc_{QQ}^{8}$ & $\erro{0.0586}{}{+27\%}{-25\%}$ & \multicolumn{2}{c\mysep|}{$\erro{0.125}{}{+10\%}{-11\%}$} & $\erro{0.00628}{}{+13\%}{-16\%}$ & $\erro{0.0133}{}{+7\%}{-5\%}$\tabularnewline
$\cc_{Qt}^{8}$ & $\erro{0.0583}{}{+27\%}{-25\%}$ & \multicolumn{2}{c\mysep|}{$\err{\shortminus0.107}{6}{+40\%}{-33\%}$} & $\erro{0.00619}{}{+13\%}{-16\%}$ & $\erro{0.0118}{}{+8\%}{-5\%}$\tabularnewline[0.5mm]
\hline
$\cc_{QQ}^{1}$ & $[\shortminus0.11_{-18\%}^{+15\%}]$ & $\err{\shortminus0.039}{4}{+51\%}{-33\%}$ & $[\shortminus0.12_{-5\%}^{+7\%}]$ & $\erro{0.0282}{}{+13\%}{-16\%}$ & $\erro{0.0651}{}{+5\%}{-6\%}$\tabularnewline
$\cc_{Qt}^{1}$ & $[\shortminus0.068_{-18\%}^{+16\%}]$ & $\erro{\shortminus2.51}{}{+29\%}{-21\%}$ & $[\shortminus0.12_{-6\%}^{+3\%}]$ & $\erro{0.0283}{}{+13\%}{-16\%}$ & $\erro{0.066}{}{+5\%}{-6\%}$\tabularnewline[0.5mm]
\cline{3-4}
$\cc_{tt}^{1}$ & $\times$ & \multicolumn{2}{c\mysep|}{$\erro{0.215}{}{+23\%}{-18\%}$} & $\times$ & $\times$\tabularnewline
\hline
\end{tabular}
}
\caption{Four-fermion contributions [pb] to top-quark pair production, at linear and quadratic levels, LO and NLO, including QCD scale uncertainties, for the LHC $\sqrt{s}=13\,$TeV and $c_i/\Lambda^2=1\,\text{TeV}^{-2}$.
The two-light two-heavy color-singlet operators (second block) only interfere at NLO with the leading QCD contribution.
The numbers in square brackets correspond to the interference with the EW contribution. 
The operators in the third block involve only third-generation quarks.
Non-vanishing contributions at $\mathcal{O}(\Lambda^{-2})$ and LO from these operators can arise through the $b\bar b$ initial state.
The SM NLO QCD cross-section is $744^{+12\%}_{-12\%}\,$pb.
}
\label{tab:ttbar}
\end{table}
\begin{table}
\renewcommand{\arraystretch}{1.4}
\centering
{\scriptsize
\newcommand{\mysep}{@{\hspace{4mm}}}
\adjustbox{max width = \columnwidth}{%
\begin{tabular}{|c|ccc|ccc|}
\hline
\multirow{2}{*}{$c_i$}
    &\multicolumn{3}{c|}{$\mathcal{O}(\Lambda^{-2})$}
    &\multicolumn{3}{c|}{$\mathcal{O}(\Lambda^{-4})$}   \\[-1mm]
    &\multicolumn{1}{c }{LO} & NLO & $K$
    &\multicolumn{1}{c }{LO} & NLO & $K$                    \\\hline\hline
$\cc_{QQ}^8$ & $\erro{0.081}{}{+55\%}{-33\%}$ $[-0.358]$ & $\erro{0.090}{}{+4\%}{-11\%}$  
& 1.1
& $\erro{0.115}{}{+46\%}{-29\%}$  & $\erro{0.158}{}{+4\%}{-11\%}$ 
& 1.37\\
$\cc_{Qt}^8$ & $\erro{0.274}{}{+54\%}{-33\%}$   $[-0.639]$      &  $\erro{0.311}{}{+5\%}{-10\%}$
& 1.14
& $\erro{0.342}{}{+46\%}{-29\%}$   & $\erro{0.378}{}{+4\%}{-13\%}$
& 1.10    \\[0.5mm]
\hline
$\cc_{QQ}^1$ & $\erro{0.242}{}{+55\%}{-33\%}$  $[-1.07]$  & $\err{0.24}{3}{+3\%}{-18\%}$ 
& 0.99
& $\erro{1.039}{}{+47\%}{-29\%}$  & $\erro{1.41}{}{+4\%}{-11\%}$ 
& 1.36 \\
$\cc_{Qt}^1$ & $\erro{-0.0098(10)}{}{+38\%}{-33\%}$ $[0.862]$  & $\erro{-0.019(9)}{}{+63\%}{-27\%}$
& 1.9
& $\erro{1.406}{}{+46\%}{-30\%}$       & $\erro{1.86}{}{+4\%}{-10\%}$  
& 1.32 \\
$\cc_{tt}^1$ & $\erro{0.483}{}{+55\%}{-33\%}$  $[-1.86]$  & $\err{0.53}{8}{+3\%}{-10\%}$ 
& 1.10
& $\erro{4.154}{}{+47\%}{-29\%}$  & $\erro{5.61}{}{+4\%}{-11\%}$ 
& 1.35 \\
\hline
\end{tabular}}}
\caption{Third-generation four-fermion operator contributions [fb] to $t\bar t t\bar t$ production at the LHC $\sqrt{s}=13\,$TeV, with $K$-factors ($\equiv\sigma_{\rm NLO}/\sigma_{\rm LO}$).
The LO interferences with SM amplitudes of order $\alpha_S\alpha_\text{EW}$ and $\alpha_Sy_t^2$ are indicated in square brackets.
The SM NLO QCD cross-section is $11.1^{+25\%}_{-25\%}\,$fb ($K=1.84$).
}
\label{tab:4tops}
\end{table}
Another interesting possibility that opens up at NLO is to probe the third-generation four-quark operators (last five rows in \autoref{tab:ttbar}), using $t/b$-loop induced effects in $gg/q\bar q$-initiated channels.
They are otherwise mainly constrained by $t\bar tb\bar b$ and $t\bar tt\bar t$ production.
Operators involving doublets already contribute at LO in the $b\bar b \to t \bar t$ channel, but suppressed by the $b$-quark luminosity.
Remarkably, the linear NLO contributions span two orders of magnitude.
Cancellations occur between partonic channels and phase-space regions for all coefficients other than $\cc{^1_{Qt}}$ and lead to an order-of-magnitude suppression for $\cc{^1_{QQ}}$.
As shown in \autoref{fig:mtt}, the contributions from color-singlet $\cc{^1_{QQ}}$ and $\cc{^1_{tt}}$ change sign around $m(t\bar{t})=400$--$450\,$GeV.
Their quark- and gluon-channel components also have opposite signs across the whole invariant-mass distribution.
Partial cancellations also occur, for $\cc{^8_{QQ}}$, between quark and gluon channels above $m(t\bar{t})\simeq 400\,$GeV and, for $\cc^8_{Qt}$, between the $b\bar{b}$ channel and others.
Although these NLO dependencies are small, they could potentially be isolated by exploiting differential distributions in $t\bar{t}$ final states.
It is instructive to compare these sensitivities to those of $t\bar t t\bar t$ production, for which evidence has recently been obtained at the LHC~\cite{Sirunyan:2019wxt,Aad:2020klt}.
The $t\bar t t\bar t$ dependencies are computed for the first time at NLO and provided in \autoref{tab:4tops} together with their $K$-factors (NLO over LO rates).
To facilitate a comparison with \autoref{tab:ttbar}, we define operator coefficients at $m_t$.
QCD renormalization and factorization scales are instead fixed to $2m_t$.
The $K$-factors of linear dependencies are close to one, except for $\cc_{Qt}^1$ where NLO corrections lift strong phase-space cancellations occurring at LO.
This suppressed interference for $\cc_{Qt}^1$ in $t\bar tt\bar t$ production contrasts with the relative enhancement of its loop-induced contribution to $t\bar t$ production, noted above.
\autoref{tab:4tops} also provides the LO interferences with subleading SM amplitudes of order $\alpha_S\alpha_\text{EW}$ and $\alpha_Sy_t^2$ which are actually larger than with the leading QCD ones.
Note they also have opposite signs.
At the quadratic level, the NLO enhancement factors reach about $1.3$ but remain smaller than the SM one at about $1.8$.
Without restriction on the energy scale probed, the current experimental sensitivity in $pp\to t\bar t t\bar t$ is dominated by energy-growing quadratic SMEFT contributions, especially for color-singlet operators which have smaller linear contributions and larger quadratic ones.
Individual sensitivities are then larger than in $pp\to t\bar t$.
Interesting complementarities between the two processes could however arise with improved measurements, for low-scale UV models, or in a global picture where various operators are to be disentangled.

As a second application, we consider pair ($W^+W^-$, $ZZ$, $W^\pm Z$) and triple ($W^+W^-W^\pm$, $W^+W^-Z$, $ZZW^\pm$, $ZZZ$) weak-boson production at the LHC $\sqrt{s}=13\,$TeV.
The latter process has just been observed at the LHC, opening a new window into electroweak gauge self-interactions~\cite{Sirunyan:2020cjp}.
The neutral final states can be produced via $gg$ fusion through a loop of fermions (at order $\alpha_S^2 \alpha_\text{EW}^2$ in the SM).
Novel SMEFT computations made available include that of triboson production at NLO in QCD, the dependence of four-quark contributions to $q \bar q \to VV$ not considered previously~\cite{Baglio:2017bfe,Baglio:2019uty}, and the full $gg \to W^+W^-$, $ZZ$ depedence extending the results of Ref.~\cite{Azatov:2014jga}.
The $gg$ fusion to $W^+W^-$ and $ZZ$ are sizeable at the LHC and probe Higgs as well as top-quark couplings.
On the contrary, the $gg$-induced production of three bosons is relatively small, with SM cross-sections for $gg \to ZZZ$ and $gg \to W^+W^-Z$ of about 0.5\% ($0.07$\,fb) and 5\% ($8.6\,$fb) of the corresponding $q\bar q$ channel~\cite{Hirschi:2015iia} at 13\,TeV.
\begin{figure}[t]
\centering
\includegraphics[trim=9.5mm 7mm 1cm 1.5mm,clip, width=0.9\columnwidth]{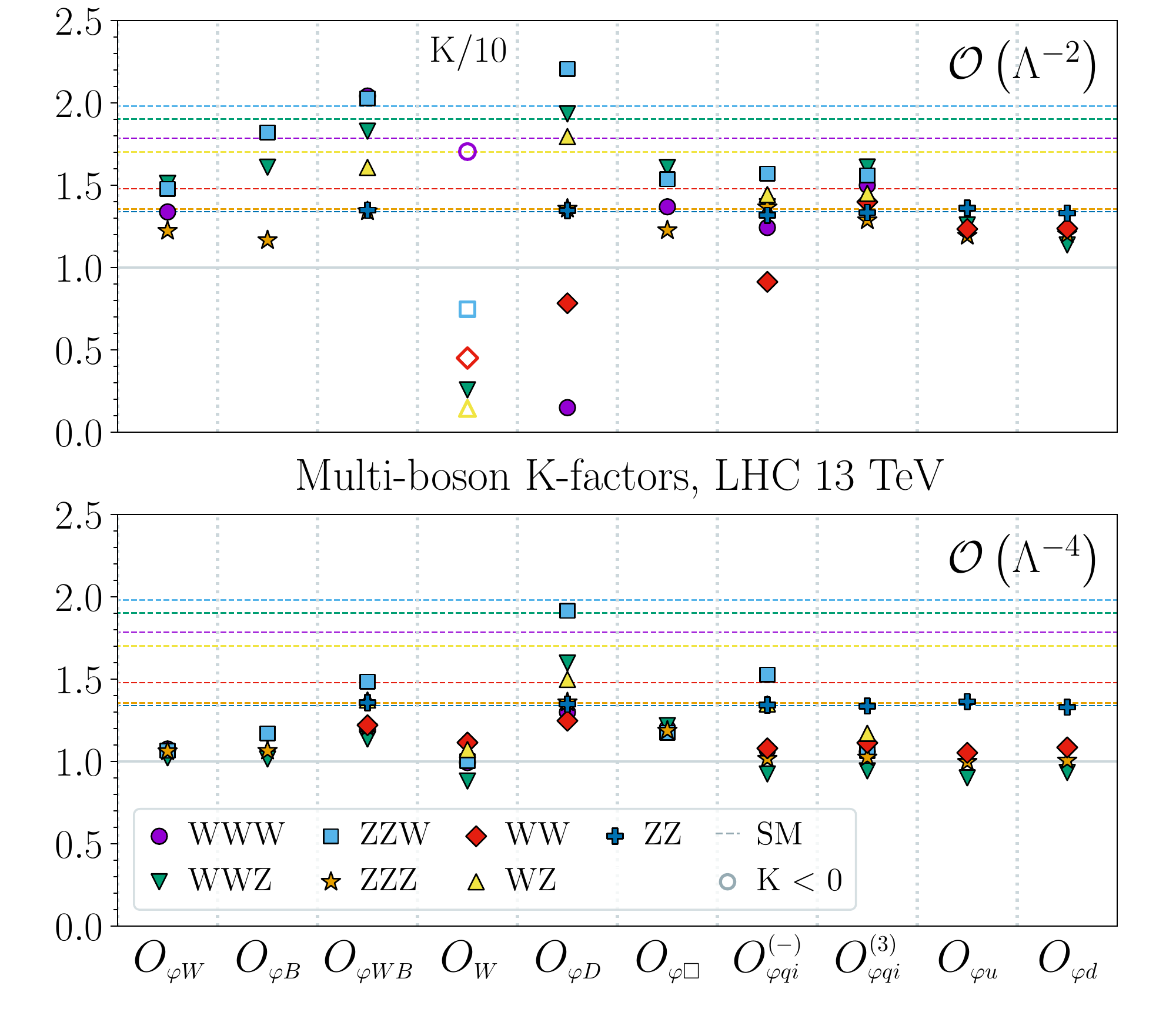}
\caption{$K$-factors (NLO/LO) of the linear $(\Lambda^{-2})$ and quadratic $(\Lambda^{-4})$ contributions to pair and triple weak-boson production at the LHC $\sqrt{s}=13\,$TeV.
Charge-conjugated final states are summed over.
$\OO_W$ values at $\mathcal{O}(\Lambda^{-2})$ are divided by $10$ and negative for empty markers.}
\label{fig:VVV}
\end{figure}
Shown in \autoref{fig:VVV} and \autoref{tab:VVV}, the $K$-factors of quark-induced channels significantly vary, not only from operator to operator, but also across processes for the same operator, and between the interference and quadratic contributions.
In general, they range between one and two.
However, for the $\OO_{W}$ operator involving three $W$ field strengths, $K$-factors at $\mathcal{O}(\Lambda^{-2})$ are extremely large and even negative, signalling that NLO corrections are lifting a suppression that occurs at LO.
It is known that the linear contribution of this operator to the inclusive diboson cross-section is very small at LO relative to the SM prediction 
($\erro{0.171}{5}{+4\%}{-5\%}\,$pb vs.\ $\erro{71.0}{1}{+6\%}{-7\%}\,$pb for $WW$) 
because of helicity selection rules~\cite{Azatov:2016sqh}, and changes sign at NLO in QCD, albeit staying below 1\% 
($\erro{-0.77}{4}{-14\%}{+16\%}\,$pb vs.\ $\erro{104}{3}{+5\%}{-5\%}\,$pb).
For $WWZ$ production, the linear LO contribution is already sizeable 
($\erro{-12.3}{5}{+1.4\%}{-1.6\%}\,$fb vs.\ $\erro{91.3}{1}{+0.0\%}{-0.5\%}\,$fb) 
and becomes larger at NLO 
($\erro{-32.0}{1}{+12\%}{-9\%}\,$fb vs.\ $\erro{173.6}{4}{+8\%}{-6\%}\,$fb).
For $W^+W^+W^-$ production the linear LO contribution is tiny 
($\err{0.4}{2}{+8\%}{-10\%}\,$fb vs.\ $\erro{79.38}{5}{+0.1\%}{-0.6\%}\,$fb) 
but becomes significant at NLO 
($\erro{-10.8}{2}{+21\%}{-16\%}\,$fb vs.\ $\erro{142.8}{9}{+7\%}{-5\%}\,$fb).
These results suggest that, in addition to spin correlation observables in $VV$~\cite{Azatov:2017kzw,Panico:2017frx}, the rates of triple-vector-boson production could help bound the $\OO_{W}$ operator.
We defer further discussions of the loop and NLO effects in multi-boson final states to a dedicated publication.
\begin{figure}[t]
\centering
\includegraphics[width=\columnwidth]{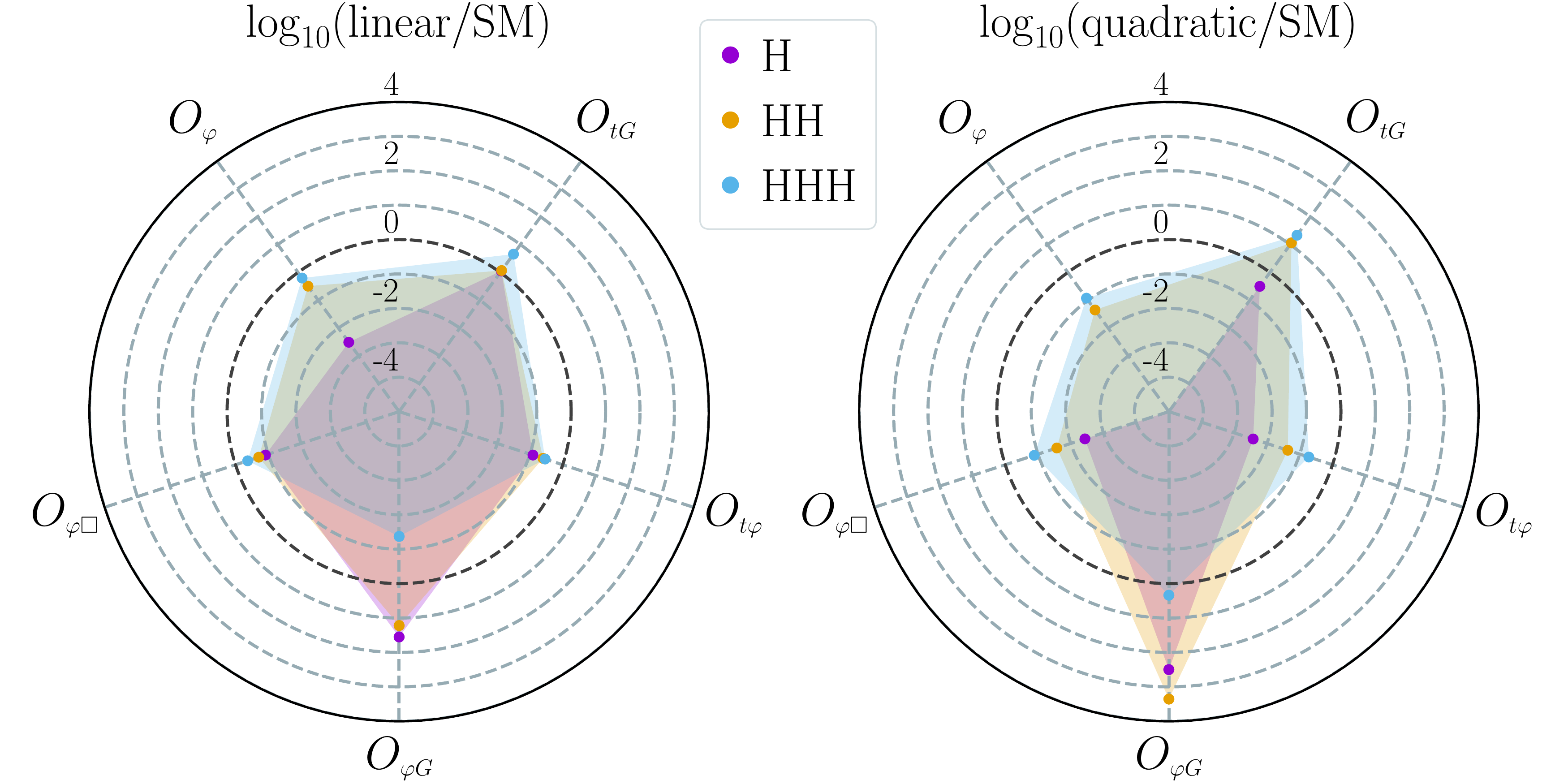}
\caption{Linear and quadratic contributions of the five relevant operators to $H$, $HH$, and $HHH$ production at a future 100\,TeV $pp$ collider, normalised by the corresponding SM predictions, for $c_i/\Lambda^2=1\,\text{TeV}^{-2}$.
}
\label{fig:radar}
\end{figure}

As a third application, we show in \autoref{fig:radar} and \autoref{tab:hhh} the sensitivity of the loop-induced Higgs production processes $gg \to H$, $HH$ and $HHH$ to various SMEFT operators in $pp$ collisions at $\sqrt{s}=100\,$TeV.
Access to all of these processes will provide the necessary information to determine trilinear and quartic terms of the Higgs potential.
Two panels display linear and quadratic contributions of $\OO_{tG}$, $\OO_{\phi G}$, $\OO_{t\phi}$, $\OO_\phi$, $\OO_{\phi\square}$ operators normalised by the SM rate.
All dependencies are calculated at one loop with SMEFT@NLO, except for the linear dependence of $gg \to H$ on $\OO_\phi$ which appears at two loops and is taken from Ref.~\cite{Degrassi:2016wml}.
The computation of SMEFT effects in $HHH$ production is presented here for the first time.
In general, the sensitivity to the various operators increases with the final state multiplicity, partially compensating the loss in statistical power due to the decreasing rates.
The only exception is $O_{\phi G}$ whose contribution to $HHH$ is suppressed by an off-shell Higgs propagator.
The loss of statistics is reflected in the projected FCC-hh reach: 1\%, 5\% and 50\% on $H$, $HH$ and $HHH$ \cite{Abada:2019lih,Mangano:2020sao,Papaefstathiou:2019ofh}, respectively.
Though challenging, $HHH$ production might be used as a diagnostic process, should a significant $O_{\phi}$-like deviation be observed in $HH$, given its larger relative sensitivity in this parameter.

\subparagraph{Conclusions}
In this article, we have presented the automation of SMEFT computations up to one-loop accuracy, illustrated with selected phenomenological applications for the LHC and future colliders.
Providing necessary input for the extraction of operator coefficients, the implementation can readily be used in current experimental and theoretical interpretations of collider data where it opens the possibility to systematically leverage NLO accuracy, reduced theoretical uncertainties, and loop-induced sensitivities in the SMEFT. 

Several directions of further developments can be identified.
The first is to extend our implementation to the elements necessary for EW loop computations, building upon the existing automation of EW corrections in the SM~\cite{Frederix:2018nkq} and the available analytic results in the SMEFT~\cite{Hartmann:2015oia, Ghezzi:2015vva, Hartmann:2016pil, Dawson:2018pyl, Dedes:2018seb, Dawson:2018liq, Dawson:2018jlg,  Dedes:2019bew, Cullen:2019nnr, Dawson:2019clf, Martini:2019lsi, Cullen:2020zof}. 
Dedicated studies of one-loop EW effects have already appeared~\cite{Vryonidou:2018eyv, Durieux:2018ggn}.
The second is to exploit the modularity of the implementation to lift some of the working assumptions, for example by including CP-violating couplings, more general flavor structures, FCNC interactions~\cite{Degrande:2014tta, Durieux:2014xla, TopFCNC}, or higher-dimensional operators.
Finally, operator coefficients are kept at a fixed scale in the current implementation.
Work is ongoing to include their renormalization-group evolution to better describe distributions spanning a wide range of scales and to automatically determine full scale uncertainties.

\subparagraph{Acknowledgements}
We are grateful to Olivier Mattelaer and Marco Zaro for the support with the {\sc MadGraph5\-\_aMC@NLO} implementation.
C.Z.\ would like to thank Christoph Lehner for helpful discussions regarding the evanescent operators.
This work has received funding by the Marie Sk\l{}odowska-Curie MCnetITN3 (grant agreement n.~722104) and by the F.R.S.-FNRS with the EOS - be.h project n.~30820817 and MISU convention F.6001.19.
Computational resources have been provided by the CISM at UCLouvain and the C\'ECI funded by the F.R.S.-FNRS under convention 2.5020.11.
The work of G.D.\ is supported in part at the Technion by a fellowship from the Lady Davis Foundation.
C.Z.\ is supported by IHEP under Contract n.~Y7515540U1 and by National Natural Science Foundation of China under grant n.~12035008 and 12075256.

\appendix
\begin{table}[h!]
\renewcommand{\arraystretch}{1.4}
\centering
{\scriptsize
\newcommand{\mysep}{@{\hspace{4mm}}}
\adjustbox{max width = \columnwidth}{%
\begin{tabular}{|c|c|c|c|c|c|c|c|}
\hline
{$K$} &   WW &   WZ &  ZZ &   WWW & WWZ &  ZZW & ZZZ \\\hline
SM                                                             &  1.5 &  1.7 & 1.3 &   1.8 & 1.9 &  2.0 & 1.4 \\\hline
\multicolumn{8}{|c|}{$\mathcal{O}(\Lambda^{-2})$}\\\hline
$c_{\scriptscriptstyle W}$                                     & -4.5 & -1.4 & $-$ & -17 & 2.6 & -7.5 & $-$ \\
$c_{\scriptscriptstyle \varphi W}$                             &  $-$ &  $-$ & $-$ &   1.3 & 1.5 &  1.5 & 1.2 \\
$c_{\scriptscriptstyle \varphi B}$                             &  $-$ &  $-$ & $-$ &   $-$ & 1.6 &  1.8 & 1.2 \\
$c_{\scriptscriptstyle \varphi WB}$                            &  2.8 &  1.6 & 1.3 &   2.0 & 1.8 &  2.0 & 1.3 \\
$c_{\scriptscriptstyle \varphi D}\hphantom{O^i}$               &  0.8 &  1.8 & 1.3 &   0.1 & 1.9 &  2.2 & 1.4 \\
$c_{\scriptscriptstyle \varphi\Box}$                           &  $-$ &  $-$ & $-$ &   1.4 & 1.6 &  1.5 & 1.2 \\
$c_{ \scriptscriptstyle \varphi q_i}^{\scriptscriptstyle (-)}$ &  0.9 &  1.4 & 1.3 &   1.2 & 1.4 &  1.6 & 1.4 \\
$c_{ \scriptscriptstyle \varphi q_i}^{\scriptscriptstyle (3)}$ &  1.4 &  1.5 & 1.3 &   1.5 & 1.6 &  1.6 & 1.3 \\
$c_{ \scriptscriptstyle \varphi u}$                            &  1.2 &  $-$ & 1.4 &   $-$ & 1.3 &  $-$ & 1.2 \\
$c_{ \scriptscriptstyle \varphi d}$                            &  1.2 &  $-$ & 1.3 &   $-$ & 1.1 &  $-$ & 1.2 \\\hline
\multicolumn{8}{|c|}{$\mathcal{O}(\Lambda^{-4})$}\\\hline
$c_{\scriptscriptstyle W}$                                     & 1.1 & 1.1 & $-$ & 1.0 & 0.9 & 1.0 & $-$ \\
$c_{\scriptscriptstyle \varphi W}$                             & $-$ & $-$ & $-$ & 1.1 & 1.0 & 1.1 & 1.1 \\
$c_{\scriptscriptstyle \varphi B}$                             & $-$ & $-$ & $-$ & $-$ & 1.0 & 1.2 & 1.1 \\
$c_{\scriptscriptstyle \varphi WB}$                            & 1.2 & 1.4 & 1.4 & 1.2 & 1.1 & 1.5 & 1.2 \\
$c_{\scriptscriptstyle \varphi D}\hphantom{O^i}$               & 1.2 & 1.5 & 1.3 & 1.3 & 1.6 & 1.9 & 1.4 \\
$c_{\scriptscriptstyle \varphi\Box}$                           & $-$ & $-$ & $-$ & 1.2 & 1.2 & 1.2 & 1.2 \\
$c_{ \scriptscriptstyle \varphi q_i}^{\scriptscriptstyle (-)}$ & 1.1 & 1.3 & 1.3 & 1.1 & 0.9 & 1.5 & 1.0 \\
$c_{ \scriptscriptstyle \varphi q_i}^{\scriptscriptstyle (3)}$ & 1.1 & 1.2 & 1.3 & 1.1 & 0.9 & 1.1 & 1.0 \\
$c_{ \scriptscriptstyle \varphi u}$                            & 1.1 & $-$ & 1.4 & $-$ & 0.9 & $-$ & 1.0 \\
$c_{ \scriptscriptstyle \varphi d}$                            & 1.1 & $-$ & 1.3 & $-$ & 0.9 & $-$ & 1.0 \\
\hline
\end{tabular}

}
}
\caption{$K$-factors for various multiboson production processes in the SM and the SMEFT as shown in \autoref{fig:VVV}. Linear and quadratic SMEFT contributions are listed by operator assuming $c_i/\Lambda^2=1$.
}
\label{tab:VVV}
\end{table}

\begin{table}[h!]
\renewcommand{\arraystretch}{1.4}
\centering
{\scriptsize
\newcommand{\mysep}{@{\hspace{4mm}}}
\adjustbox{max width = \columnwidth}{%
\begin{tabular}{|c|c|c|c|c|c|c|}
\hline
\multirow{2}{*}{{\normalsize $\frac{\sigma_{i}\phantom{,}}{\sigma_{ \scalebox{.5}{SM}}}$}}
&\multicolumn{2}{|c|}{H}
&\multicolumn{2}{|c|}{HH}
&\multicolumn{2}{|c|}{HHH}\\
& \multicolumn{1}{c }{$\mathcal{O}(\Lambda^{-2})$}
& \multicolumn{1}{c|}{$\mathcal{O}(\Lambda^{-4})$}
& \multicolumn{1}{c }{$\mathcal{O}(\Lambda^{-2})$}
& \multicolumn{1}{c|}{$\mathcal{O}(\Lambda^{-4})$}
& \multicolumn{1}{c }{$\mathcal{O}(\Lambda^{-2})$}
& \multicolumn{1}{c|}{$\mathcal{O}(\Lambda^{-4})$}\\
\hline\hline
$c_{\scriptscriptstyle \varphi G}$   &  36     &  320     & -17    &  2300   & -0.042  &  2.2   \\ \hline
$c_{\scriptscriptstyle t\varphi }$   & -0.12   &  0.0038  &  0.25  &  0.043  &  0.29   &  0.19  \\ \hline
$c_{\scriptscriptstyle tG }$         &  1.1    &  0.31    & -1.2   &  13     & -4.5    &  22    \\ \hline
$c_{\scriptscriptstyle \varphi }$    & -0.0031 &  $-$     &  0.32  &  0.045  &  0.63   &  0.12  \\ \hline
$c_{\scriptscriptstyle \varphi\Box}$ &  0.12   &  0.0037  &  0.20  &  0.027  & -0.42   &  0.13  \\ \hline
\end{tabular}
}}
\caption{
Numerical values for SMEFT contributions to $H$, $HH$, and $HHH$ shown in \autoref{fig:radar}. 
Entires are normalised by the corresponding SM predictions of 340~pb, 1.0~pb and 37~fb, respectively.
}
\label{tab:hhh}
\end{table}

\bibliography{refs.bib}

\begin{thebibliography}{70}%
\makeatletter
\providecommand \@ifxundefined [1]{%
 \@ifx{#1\undefined}
}%
\providecommand \@ifnum [1]{%
 \ifnum #1\expandafter \@firstoftwo
 \else \expandafter \@secondoftwo
 \fi
}%
\providecommand \@ifx [1]{%
 \ifx #1\expandafter \@firstoftwo
 \else \expandafter \@secondoftwo
 \fi
}%
\providecommand \natexlab [1]{#1}%
\providecommand \enquote  [1]{``#1''}%
\providecommand \bibnamefont  [1]{#1}%
\providecommand \bibfnamefont [1]{#1}%
\providecommand \citenamefont [1]{#1}%
\providecommand \href@noop [0]{\@secondoftwo}%
\providecommand \href [0]{\begingroup \@sanitize@url \@href}%
\providecommand \@href[1]{\@@startlink{#1}\@@href}%
\providecommand \@@href[1]{\endgroup#1\@@endlink}%
\providecommand \@sanitize@url [0]{\catcode `\\12\catcode `\$12\catcode
  `\&12\catcode `\#12\catcode `\^12\catcode `\_12\catcode `\%12\relax}%
\providecommand \@@startlink[1]{}%
\providecommand \@@endlink[0]{}%
\providecommand \url  [0]{\begingroup\@sanitize@url \@url }%
\providecommand \@url [1]{\endgroup\@href {#1}{\urlprefix }}%
\providecommand \urlprefix  [0]{URL }%
\providecommand \Eprint [0]{\href }%
\providecommand \doibase [0]{http://dx.doi.org/}%
\providecommand \selectlanguage [0]{\@gobble}%
\providecommand \bibinfo  [0]{\@secondoftwo}%
\providecommand \bibfield  [0]{\@secondoftwo}%
\providecommand \translation [1]{[#1]}%
\providecommand \BibitemOpen [0]{}%
\providecommand \bibitemStop [0]{}%
\providecommand \bibitemNoStop [0]{.\EOS\space}%
\providecommand \EOS [0]{\spacefactor3000\relax}%
\providecommand \BibitemShut  [1]{\csname bibitem#1\endcsname}%
\let\auto@bib@innerbib\@empty
\bibitem [{\citenamefont {Weinberg}(1979)}]{Weinberg:1978kz}%
  \BibitemOpen
  \bibfield  {author} {\bibinfo {author} {\bibfnamefont {Steven}\ \bibnamefont
  {Weinberg}},\ }\bibfield  {title} {\enquote {\bibinfo {title}
  {{Phenomenological Lagrangians}},}\ }\href {\doibase
  10.1016/0378-4371(79)90223-1} {\bibfield  {journal} {\bibinfo  {journal}
  {Physica}\ }\textbf {\bibinfo {volume} {A96}},\ \bibinfo {pages} {327--340}
  (\bibinfo {year} {1979})}\BibitemShut {NoStop}%
\bibitem [{\citenamefont {Buchmuller}\ and\ \citenamefont
  {Wyler}(1986)}]{Buchmuller:1985jz}%
  \BibitemOpen
  \bibfield  {author} {\bibinfo {author} {\bibfnamefont {W.}~\bibnamefont
  {Buchmuller}}\ and\ \bibinfo {author} {\bibfnamefont {D.}~\bibnamefont
  {Wyler}},\ }\bibfield  {title} {\enquote {\bibinfo {title} {{Effective
  Lagrangian Analysis of New Interactions and Flavor Conservation}},}\ }\href
  {\doibase 10.1016/0550-3213(86)90262-2} {\bibfield  {journal} {\bibinfo
  {journal} {Nucl. Phys.}\ }\textbf {\bibinfo {volume} {B268}},\ \bibinfo
  {pages} {621--653} (\bibinfo {year} {1986})}\BibitemShut {NoStop}%
\bibitem [{\citenamefont {Leung}\ \emph {et~al.}(1986)\citenamefont {Leung},
  \citenamefont {Love},\ and\ \citenamefont {Rao}}]{Leung:1984ni}%
  \BibitemOpen
  \bibfield  {author} {\bibinfo {author} {\bibfnamefont {Chung~Ngoc}\
  \bibnamefont {Leung}}, \bibinfo {author} {\bibfnamefont {S.~T.}\ \bibnamefont
  {Love}}, \ and\ \bibinfo {author} {\bibfnamefont {S.}~\bibnamefont {Rao}},\
  }\bibfield  {title} {\enquote {\bibinfo {title} {{Low-Energy Manifestations
  of a New Interaction Scale: Operator Analysis}},}\ }\href {\doibase
  10.1007/BF01588041} {\bibfield  {journal} {\bibinfo  {journal} {Z. Phys.}\
  }\textbf {\bibinfo {volume} {C31}},\ \bibinfo {pages} {433} (\bibinfo {year}
  {1986})}\BibitemShut {NoStop}%
\bibitem [{\citenamefont {Henning}\ \emph {et~al.}(2017)\citenamefont
  {Henning}, \citenamefont {Lu}, \citenamefont {Melia},\ and\ \citenamefont
  {Murayama}}]{Henning:2015alf}%
  \BibitemOpen
  \bibfield  {author} {\bibinfo {author} {\bibfnamefont {Brian}\ \bibnamefont
  {Henning}}, \bibinfo {author} {\bibfnamefont {Xiaochuan}\ \bibnamefont {Lu}},
  \bibinfo {author} {\bibfnamefont {Tom}\ \bibnamefont {Melia}}, \ and\
  \bibinfo {author} {\bibfnamefont {Hitoshi}\ \bibnamefont {Murayama}},\
  }\bibfield  {title} {\enquote {\bibinfo {title} {{2, 84, 30, 993, 560, 15456,
  11962, 261485, ...: Higher dimension operators in the SM EFT}},}\ }\href
  {\doibase 10.1007/JHEP09(2019)019, 10.1007/JHEP08(2017)016} {\bibfield
  {journal} {\bibinfo  {journal} {JHEP}\ }\textbf {\bibinfo {volume} {08}},\
  \bibinfo {pages} {016} (\bibinfo {year} {2017})},\ \bibinfo {note} {[Erratum:
  JHEP09,019(2019)]},\ \Eprint
  {http://arxiv.org/abs/1512.03433}{arXiv:1512.03433 [hep-ph]}\BibitemShut
  {NoStop}%
\bibitem [{\citenamefont {Grzadkowski}\ \emph {et~al.}(2010)\citenamefont
  {Grzadkowski}, \citenamefont {Iskrzynski}, \citenamefont {Misiak},\ and\
  \citenamefont {Rosiek}}]{Grzadkowski:2010es}%
  \BibitemOpen
  \bibfield  {author} {\bibinfo {author} {\bibfnamefont {B.}~\bibnamefont
  {Grzadkowski}}, \bibinfo {author} {\bibfnamefont {M.}~\bibnamefont
  {Iskrzynski}}, \bibinfo {author} {\bibfnamefont {M.}~\bibnamefont {Misiak}},
  \ and\ \bibinfo {author} {\bibfnamefont {J.}~\bibnamefont {Rosiek}},\
  }\bibfield  {title} {\enquote {\bibinfo {title} {{Dimension-Six Terms in the
  Standard Model Lagrangian}},}\ }\href {\doibase 10.1007/JHEP10(2010)085}
  {\bibfield  {journal} {\bibinfo  {journal} {JHEP}\ }\textbf {\bibinfo
  {volume} {10}},\ \bibinfo {pages} {085} (\bibinfo {year} {2010})},\ \Eprint
  {http://arxiv.org/abs/1008.4884}{arXiv:1008.4884 [hep-ph]}\BibitemShut
  {NoStop}%
\bibitem [{\citenamefont {Alloul}\ \emph {et~al.}(2014)\citenamefont {Alloul},
  \citenamefont {Christensen}, \citenamefont {Degrande}, \citenamefont {Duhr},\
  and\ \citenamefont {Fuks}}]{Alloul:2013bka}%
  \BibitemOpen
  \bibfield  {author} {\bibinfo {author} {\bibfnamefont {Adam}\ \bibnamefont
  {Alloul}}, \bibinfo {author} {\bibfnamefont {Neil~D.}\ \bibnamefont
  {Christensen}}, \bibinfo {author} {\bibfnamefont {Céline}\ \bibnamefont
  {Degrande}}, \bibinfo {author} {\bibfnamefont {Claude}\ \bibnamefont {Duhr}},
  \ and\ \bibinfo {author} {\bibfnamefont {Benjamin}\ \bibnamefont {Fuks}},\
  }\bibfield  {title} {\enquote {\bibinfo {title} {{FeynRules 2.0 - A complete
  toolbox for tree-level phenomenology}},}\ }\href {\doibase
  10.1016/j.cpc.2014.04.012} {\bibfield  {journal} {\bibinfo  {journal}
  {Comput.Phys.Commun.}\ }\textbf {\bibinfo {volume} {185}},\ \bibinfo {pages}
  {2250--2300} (\bibinfo {year} {2014})},\ \Eprint
  {http://arxiv.org/abs/1310.1921}{arXiv:1310.1921 [hep-ph]}\BibitemShut
  {NoStop}%
\bibitem [{\citenamefont {Degrande}\ \emph {et~al.}()\citenamefont {Degrande},
  \citenamefont {Durieux}, \citenamefont {Maltoni}, \citenamefont {Mimasu},
  \citenamefont {Vryonidou},\ and\ \citenamefont {Zhang}}]{smeft:2020}%
  \BibitemOpen
  \bibfield  {author} {\bibinfo {author} {\bibfnamefont {C\'eline}\
  \bibnamefont {Degrande}}, \bibinfo {author} {\bibfnamefont {Gauthier}\
  \bibnamefont {Durieux}}, \bibinfo {author} {\bibfnamefont {Fabio}\
  \bibnamefont {Maltoni}}, \bibinfo {author} {\bibfnamefont {Ken}\ \bibnamefont
  {Mimasu}}, \bibinfo {author} {\bibfnamefont {Eleni}\ \bibnamefont
  {Vryonidou}}, \ and\ \bibinfo {author} {\bibfnamefont {Cen}\ \bibnamefont
  {Zhang}},\ }\bibfield  {title} {\enquote {\bibinfo {title} {{SMEFT@NLO}},}\
  }\href@noop {} {\ }\bibinfo {note}
  {\url{http://feynrules.irmp.ucl.ac.be/wiki/SMEFTatNLO}}\BibitemShut {NoStop}%
\bibitem [{\citenamefont {Aguilar-Saavedra}\ \emph {et~al.}()\citenamefont
  {Aguilar-Saavedra}, \citenamefont {Degrande}, \citenamefont {Durieux},
  \citenamefont {Maltoni}, \citenamefont {Vryonidou}, \citenamefont {Zhang}
  \emph {et~al.}}]{AguilarSaavedra:2018nen}%
  \BibitemOpen
  \bibfield  {author} {\bibinfo {author} {\bibfnamefont {J.~A.}\ \bibnamefont
  {Aguilar-Saavedra}}, \bibinfo {author} {\bibfnamefont {C.}~\bibnamefont
  {Degrande}}, \bibinfo {author} {\bibfnamefont {G.}~\bibnamefont {Durieux}},
  \bibinfo {author} {\bibfnamefont {F.}~\bibnamefont {Maltoni}}, \bibinfo
  {author} {\bibfnamefont {E.}~\bibnamefont {Vryonidou}}, \bibinfo {author}
  {\bibfnamefont {C.}~\bibnamefont {Zhang}},  \emph {et~al.},\ }\bibfield
  {title} {\enquote {\bibinfo {title} {{Interpreting top-quark LHC measurements
  in the standard-model effective field theory}},}\ }\href@noop {} {\ }\Eprint
  {http://arxiv.org/abs/1802.07237}{arXiv:1802.07237 [hep-ph]}\BibitemShut
  {NoStop}%
\bibitem [{\citenamefont {Alwall}\ \emph {et~al.}(2014)\citenamefont {Alwall},
  \citenamefont {Frederix}, \citenamefont {Frixione}, \citenamefont {Hirschi},
  \citenamefont {Maltoni} \emph {et~al.}}]{Alwall:2014hca}%
  \BibitemOpen
  \bibfield  {author} {\bibinfo {author} {\bibfnamefont {J.}~\bibnamefont
  {Alwall}}, \bibinfo {author} {\bibfnamefont {R.}~\bibnamefont {Frederix}},
  \bibinfo {author} {\bibfnamefont {S.}~\bibnamefont {Frixione}}, \bibinfo
  {author} {\bibfnamefont {V.}~\bibnamefont {Hirschi}}, \bibinfo {author}
  {\bibfnamefont {F.}~\bibnamefont {Maltoni}},  \emph {et~al.},\ }\bibfield
  {title} {\enquote {\bibinfo {title} {{The automated computation of tree-level
  and next-to-leading order differential cross sections, and their matching to
  parton shower simulations}},}\ }\href {\doibase 10.1007/JHEP07(2014)079}
  {\bibfield  {journal} {\bibinfo  {journal} {JHEP}\ }\textbf {\bibinfo
  {volume} {1407}},\ \bibinfo {pages} {079} (\bibinfo {year} {2014})},\ \Eprint
  {http://arxiv.org/abs/1405.0301}{arXiv:1405.0301 [hep-ph]}\BibitemShut
  {NoStop}%
\bibitem [{\citenamefont {Brivio}\ \emph {et~al.}(2017)\citenamefont {Brivio},
  \citenamefont {Jiang},\ and\ \citenamefont {Trott}}]{Brivio:2017btx}%
  \BibitemOpen
  \bibfield  {author} {\bibinfo {author} {\bibfnamefont {Ilaria}\ \bibnamefont
  {Brivio}}, \bibinfo {author} {\bibfnamefont {Yun}\ \bibnamefont {Jiang}}, \
  and\ \bibinfo {author} {\bibfnamefont {Michael}\ \bibnamefont {Trott}},\
  }\bibfield  {title} {\enquote {\bibinfo {title} {{The SMEFTsim package,
  theory and tools}},}\ }\href {\doibase 10.1007/JHEP12(2017)070} {\bibfield
  {journal} {\bibinfo  {journal} {JHEP}\ }\textbf {\bibinfo {volume} {12}},\
  \bibinfo {pages} {070} (\bibinfo {year} {2017})},\ \Eprint
  {http://arxiv.org/abs/1709.06492}{arXiv:1709.06492 [hep-ph]}\BibitemShut
  {NoStop}%
\bibitem [{\citenamefont {Durieux}\ \emph {et~al.}()\citenamefont {Durieux},
  \citenamefont {Brivio}, \citenamefont {Maltoni}, \citenamefont {Trott} \emph
  {et~al.}}]{Durieux:2019lnv}%
  \BibitemOpen
  \bibfield  {author} {\bibinfo {author} {\bibfnamefont {Gauthier}\
  \bibnamefont {Durieux}}, \bibinfo {author} {\bibfnamefont {Ilaria}\
  \bibnamefont {Brivio}}, \bibinfo {author} {\bibfnamefont {Fabio}\
  \bibnamefont {Maltoni}}, \bibinfo {author} {\bibfnamefont {Michael}\
  \bibnamefont {Trott}},  \emph {et~al.},\ }\bibfield  {title} {\enquote
  {\bibinfo {title} {{Proposal for the validation of Monte Carlo
  implementations of the standard model effective field theory}},}\ }\href@noop
  {} {\ }\Eprint {http://arxiv.org/abs/1906.12310}{arXiv:1906.12310
  [hep-ph]}\BibitemShut {NoStop}%
\bibitem [{\citenamefont {Ossola}\ \emph {et~al.}(2007)\citenamefont {Ossola},
  \citenamefont {Papadopoulos},\ and\ \citenamefont {Pittau}}]{Ossola:2006us}%
  \BibitemOpen
  \bibfield  {author} {\bibinfo {author} {\bibfnamefont {Giovanni}\
  \bibnamefont {Ossola}}, \bibinfo {author} {\bibfnamefont {Costas~G.}\
  \bibnamefont {Papadopoulos}}, \ and\ \bibinfo {author} {\bibfnamefont
  {Roberto}\ \bibnamefont {Pittau}},\ }\bibfield  {title} {\enquote {\bibinfo
  {title} {{Reducing full one-loop amplitudes to scalar integrals at the
  integrand level}},}\ }\href {\doibase 10.1016/j.nuclphysb.2006.11.012}
  {\bibfield  {journal} {\bibinfo  {journal} {Nucl. Phys.}\ }\textbf {\bibinfo
  {volume} {B763}},\ \bibinfo {pages} {147--169} (\bibinfo {year} {2007})},\
  \Eprint {http://arxiv.org/abs/hep-ph/0609007}{arXiv:hep-ph/0609007
  [hep-ph]}\BibitemShut {NoStop}%
\bibitem [{\citenamefont {Jenkins}\ \emph {et~al.}(2013)\citenamefont
  {Jenkins}, \citenamefont {Manohar},\ and\ \citenamefont
  {Trott}}]{Jenkins:2013zja}%
  \BibitemOpen
  \bibfield  {author} {\bibinfo {author} {\bibfnamefont {Elizabeth~E.}\
  \bibnamefont {Jenkins}}, \bibinfo {author} {\bibfnamefont {Aneesh~V.}\
  \bibnamefont {Manohar}}, \ and\ \bibinfo {author} {\bibfnamefont {Michael}\
  \bibnamefont {Trott}},\ }\bibfield  {title} {\enquote {\bibinfo {title}
  {{Renormalization Group Evolution of the Standard Model Dimension Six
  Operators I: Formalism and lambda Dependence}},}\ }\href {\doibase
  10.1007/JHEP10(2013)087} {\bibfield  {journal} {\bibinfo  {journal} {JHEP}\
  }\textbf {\bibinfo {volume} {10}},\ \bibinfo {pages} {087} (\bibinfo {year}
  {2013})},\ \Eprint {http://arxiv.org/abs/1308.2627}{arXiv:1308.2627
  [hep-ph]}\BibitemShut {NoStop}%
\bibitem [{\citenamefont {Jenkins}\ \emph {et~al.}(2014)\citenamefont
  {Jenkins}, \citenamefont {Manohar},\ and\ \citenamefont
  {Trott}}]{Jenkins:2013wua}%
  \BibitemOpen
  \bibfield  {author} {\bibinfo {author} {\bibfnamefont {Elizabeth~E.}\
  \bibnamefont {Jenkins}}, \bibinfo {author} {\bibfnamefont {Aneesh~V.}\
  \bibnamefont {Manohar}}, \ and\ \bibinfo {author} {\bibfnamefont {Michael}\
  \bibnamefont {Trott}},\ }\bibfield  {title} {\enquote {\bibinfo {title}
  {{Renormalization Group Evolution of the Standard Model Dimension Six
  Operators II: Yukawa Dependence}},}\ }\href {\doibase
  10.1007/JHEP01(2014)035} {\bibfield  {journal} {\bibinfo  {journal} {JHEP}\
  }\textbf {\bibinfo {volume} {01}},\ \bibinfo {pages} {035} (\bibinfo {year}
  {2014})},\ \Eprint {http://arxiv.org/abs/1310.4838}{arXiv:1310.4838
  [hep-ph]}\BibitemShut {NoStop}%
\bibitem [{\citenamefont {Alonso}\ \emph {et~al.}(2014)\citenamefont {Alonso},
  \citenamefont {Jenkins}, \citenamefont {Manohar},\ and\ \citenamefont
  {Trott}}]{Alonso:2013hga}%
  \BibitemOpen
  \bibfield  {author} {\bibinfo {author} {\bibfnamefont {Rodrigo}\ \bibnamefont
  {Alonso}}, \bibinfo {author} {\bibfnamefont {Elizabeth~E.}\ \bibnamefont
  {Jenkins}}, \bibinfo {author} {\bibfnamefont {Aneesh~V.}\ \bibnamefont
  {Manohar}}, \ and\ \bibinfo {author} {\bibfnamefont {Michael}\ \bibnamefont
  {Trott}},\ }\bibfield  {title} {\enquote {\bibinfo {title} {{Renormalization
  Group Evolution of the Standard Model Dimension Six Operators III: Gauge
  Coupling Dependence and Phenomenology}},}\ }\href {\doibase
  10.1007/JHEP04(2014)159} {\bibfield  {journal} {\bibinfo  {journal} {JHEP}\
  }\textbf {\bibinfo {volume} {04}},\ \bibinfo {pages} {159} (\bibinfo {year}
  {2014})},\ \Eprint {http://arxiv.org/abs/1312.2014}{arXiv:1312.2014
  [hep-ph]}\BibitemShut {NoStop}%
\bibitem [{\citenamefont {Degrande}(2015)}]{Degrande:2014vpa}%
  \BibitemOpen
  \bibfield  {author} {\bibinfo {author} {\bibfnamefont {Celine}\ \bibnamefont
  {Degrande}},\ }\bibfield  {title} {\enquote {\bibinfo {title} {{Automatic
  evaluation of UV and R2 terms for beyond the Standard Model Lagrangians: a
  proof-of-principle}},}\ }\href {\doibase 10.1016/j.cpc.2015.08.015}
  {\bibfield  {journal} {\bibinfo  {journal} {Comput. Phys. Commun.}\ }\textbf
  {\bibinfo {volume} {197}},\ \bibinfo {pages} {239--262} (\bibinfo {year}
  {2015})},\ \Eprint {http://arxiv.org/abs/1406.3030}{arXiv:1406.3030
  [hep-ph]}\BibitemShut {NoStop}%
\bibitem [{\citenamefont {Bardeen}\ and\ \citenamefont
  {Zumino}(1984)}]{Bardeen:1984pm}%
  \BibitemOpen
  \bibfield  {author} {\bibinfo {author} {\bibfnamefont {William~A.}\
  \bibnamefont {Bardeen}}\ and\ \bibinfo {author} {\bibfnamefont {Bruno}\
  \bibnamefont {Zumino}},\ }\bibfield  {title} {\enquote {\bibinfo {title}
  {{Consistent and Covariant Anomalies in Gauge and Gravitational Theories}},}\
  }\href {\doibase 10.1016/0550-3213(84)90322-5} {\bibfield  {journal}
  {\bibinfo  {journal} {Nucl. Phys.}\ }\textbf {\bibinfo {volume} {B244}},\
  \bibinfo {pages} {421--453} (\bibinfo {year} {1984})}\BibitemShut {NoStop}%
\bibitem [{\citenamefont {Fox}\ \emph {et~al.}(2018)\citenamefont {Fox},
  \citenamefont {Low},\ and\ \citenamefont {Zhang}}]{Fox:2018ldq}%
  \BibitemOpen
  \bibfield  {author} {\bibinfo {author} {\bibfnamefont {Patrick~J.}\
  \bibnamefont {Fox}}, \bibinfo {author} {\bibfnamefont {Ian}\ \bibnamefont
  {Low}}, \ and\ \bibinfo {author} {\bibfnamefont {Yue}\ \bibnamefont
  {Zhang}},\ }\bibfield  {title} {\enquote {\bibinfo {title} {{Top-philic $Z'$
  forces at the LHC}},}\ }\href {\doibase 10.1007/JHEP03(2018)074} {\bibfield
  {journal} {\bibinfo  {journal} {JHEP}\ }\textbf {\bibinfo {volume} {03}},\
  \bibinfo {pages} {074} (\bibinfo {year} {2018})},\ \Eprint
  {http://arxiv.org/abs/1801.03505}{arXiv:1801.03505 [hep-ph]}\BibitemShut
  {NoStop}%
\bibitem [{\citenamefont {Kreimer}(1990)}]{Kreimer:1989ke}%
  \BibitemOpen
  \bibfield  {author} {\bibinfo {author} {\bibfnamefont {Dirk}\ \bibnamefont
  {Kreimer}},\ }\bibfield  {title} {\enquote {\bibinfo {title} {{The $\gamma_5$
  Problem and Anomalies: A Clifford Algebra Approach}},}\ }\href {\doibase
  10.1016/0370-2693(90)90461-E} {\bibfield  {journal} {\bibinfo  {journal}
  {Phys. Lett. B}\ }\textbf {\bibinfo {volume} {237}},\ \bibinfo {pages}
  {59--62} (\bibinfo {year} {1990})}\BibitemShut {NoStop}%
\bibitem [{\citenamefont {Korner}\ \emph {et~al.}(1992)\citenamefont {Korner},
  \citenamefont {Kreimer},\ and\ \citenamefont {Schilcher}}]{Korner:1991sx}%
  \BibitemOpen
  \bibfield  {author} {\bibinfo {author} {\bibfnamefont {J.~G.}\ \bibnamefont
  {Korner}}, \bibinfo {author} {\bibfnamefont {D.}~\bibnamefont {Kreimer}}, \
  and\ \bibinfo {author} {\bibfnamefont {K.}~\bibnamefont {Schilcher}},\
  }\bibfield  {title} {\enquote {\bibinfo {title} {{A Practicable $\gamma_5$
  scheme in dimensional regularization}},}\ }\href {\doibase
  10.1007/BF01559471} {\bibfield  {journal} {\bibinfo  {journal} {Z. Phys. C}\
  }\textbf {\bibinfo {volume} {54}},\ \bibinfo {pages} {503--512} (\bibinfo
  {year} {1992})}\BibitemShut {NoStop}%
\bibitem [{\citenamefont {Kreimer}(1993)}]{Kreimer:1993bh}%
  \BibitemOpen
  \bibfield  {author} {\bibinfo {author} {\bibfnamefont {Dirk}\ \bibnamefont
  {Kreimer}},\ }\bibfield  {title} {\enquote {\bibinfo {title} {{The Role of
  $\gamma_5$ in dimensional regularization}},}\ }\href@noop {} {\  (\bibinfo
  {year} {1993})},\ \Eprint
  {http://arxiv.org/abs/hep-ph/9401354}{arXiv:hep-ph/9401354}\BibitemShut
  {NoStop}%
\bibitem [{\citenamefont {Dugan}\ and\ \citenamefont
  {Grinstein}(1991)}]{Dugan:1990df}%
  \BibitemOpen
  \bibfield  {author} {\bibinfo {author} {\bibfnamefont {Michael~J.}\
  \bibnamefont {Dugan}}\ and\ \bibinfo {author} {\bibfnamefont {Benjamin}\
  \bibnamefont {Grinstein}},\ }\bibfield  {title} {\enquote {\bibinfo {title}
  {{On the vanishing of evanescent operators}},}\ }\href {\doibase
  10.1016/0370-2693(91)90680-O} {\bibfield  {journal} {\bibinfo  {journal}
  {Phys. Lett.}\ }\textbf {\bibinfo {volume} {B256}},\ \bibinfo {pages}
  {239--244} (\bibinfo {year} {1991})}\BibitemShut {NoStop}%
\bibitem [{\citenamefont {Herrlich}\ and\ \citenamefont
  {Nierste}(1995)}]{Herrlich:1994kh}%
  \BibitemOpen
  \bibfield  {author} {\bibinfo {author} {\bibfnamefont {Stefan}\ \bibnamefont
  {Herrlich}}\ and\ \bibinfo {author} {\bibfnamefont {Ulrich}\ \bibnamefont
  {Nierste}},\ }\bibfield  {title} {\enquote {\bibinfo {title} {{Evanescent
  operators, scheme dependences and double insertions}},}\ }\href {\doibase
  10.1016/0550-3213(95)00474-7} {\bibfield  {journal} {\bibinfo  {journal}
  {Nucl. Phys.}\ }\textbf {\bibinfo {volume} {B455}},\ \bibinfo {pages}
  {39--58} (\bibinfo {year} {1995})},\ \Eprint
  {http://arxiv.org/abs/hep-ph/9412375}{arXiv:hep-ph/9412375
  [hep-ph]}\BibitemShut {NoStop}%
\bibitem [{\citenamefont {Buras}\ and\ \citenamefont
  {Weisz}(1990)}]{Buras:1989xd}%
  \BibitemOpen
  \bibfield  {author} {\bibinfo {author} {\bibfnamefont {Andrzej~J.}\
  \bibnamefont {Buras}}\ and\ \bibinfo {author} {\bibfnamefont {Peter~H.}\
  \bibnamefont {Weisz}},\ }\bibfield  {title} {\enquote {\bibinfo {title} {{QCD
  Nonleading Corrections to Weak Decays in Dimensional Regularization and 't
  Hooft-Veltman Schemes}},}\ }\href {\doibase 10.1016/0550-3213(90)90223-Z}
  {\bibfield  {journal} {\bibinfo  {journal} {Nucl. Phys.}\ }\textbf {\bibinfo
  {volume} {B333}},\ \bibinfo {pages} {66--99} (\bibinfo {year}
  {1990})}\BibitemShut {NoStop}%
\bibitem [{\citenamefont {Gao}\ \emph {et~al.}(2011)\citenamefont {Gao},
  \citenamefont {Li}, \citenamefont {Wang}, \citenamefont {Zhu},\ and\
  \citenamefont {Yuan}}]{Gao:2011ha}%
  \BibitemOpen
  \bibfield  {author} {\bibinfo {author} {\bibfnamefont {Jun}\ \bibnamefont
  {Gao}}, \bibinfo {author} {\bibfnamefont {Chong~Sheng}\ \bibnamefont {Li}},
  \bibinfo {author} {\bibfnamefont {Jian}\ \bibnamefont {Wang}}, \bibinfo
  {author} {\bibfnamefont {Hua~Xing}\ \bibnamefont {Zhu}}, \ and\ \bibinfo
  {author} {\bibfnamefont {C.~P.}\ \bibnamefont {Yuan}},\ }\bibfield  {title}
  {\enquote {\bibinfo {title} {{Next-to-leading QCD effect to the quark
  compositeness search at the LHC}},}\ }\href {\doibase
  10.1103/PhysRevLett.106.142001} {\bibfield  {journal} {\bibinfo  {journal}
  {Phys. Rev. Lett.}\ }\textbf {\bibinfo {volume} {106}},\ \bibinfo {pages}
  {142001} (\bibinfo {year} {2011})},\ \Eprint
  {http://arxiv.org/abs/1101.4611}{arXiv:1101.4611 [hep-ph]}\BibitemShut
  {NoStop}%
\bibitem [{\citenamefont {Shao}\ \emph {et~al.}(2011)\citenamefont {Shao},
  \citenamefont {Li}, \citenamefont {Wang}, \citenamefont {Gao}, \citenamefont
  {Zhang},\ and\ \citenamefont {Zhu}}]{Shao:2011wa}%
  \BibitemOpen
  \bibfield  {author} {\bibinfo {author} {\bibfnamefont {Ding~Yu}\ \bibnamefont
  {Shao}}, \bibinfo {author} {\bibfnamefont {Chong~Sheng}\ \bibnamefont {Li}},
  \bibinfo {author} {\bibfnamefont {Jian}\ \bibnamefont {Wang}}, \bibinfo
  {author} {\bibfnamefont {Jun}\ \bibnamefont {Gao}}, \bibinfo {author}
  {\bibfnamefont {Hao}\ \bibnamefont {Zhang}}, \ and\ \bibinfo {author}
  {\bibfnamefont {Hua~Xing}\ \bibnamefont {Zhu}},\ }\bibfield  {title}
  {\enquote {\bibinfo {title} {{Model independent analysis of top quark
  forward-backward asymmetry at the Tevatron up to
  $\mathcal{O}(\alpha_S^2/\Lambda^2)$}},}\ }\href {\doibase
  10.1103/PhysRevD.84.054016} {\bibfield  {journal} {\bibinfo  {journal} {Phys.
  Rev.}\ }\textbf {\bibinfo {volume} {D84}},\ \bibinfo {pages} {054016}
  (\bibinfo {year} {2011})},\ \Eprint
  {http://arxiv.org/abs/1107.4012}{arXiv:1107.4012 [hep-ph]}\BibitemShut
  {NoStop}%
\bibitem [{\citenamefont {Degrande}\ \emph {et~al.}(2015)\citenamefont
  {Degrande}, \citenamefont {Maltoni}, \citenamefont {Wang},\ and\
  \citenamefont {Zhang}}]{Degrande:2014tta}%
  \BibitemOpen
  \bibfield  {author} {\bibinfo {author} {\bibfnamefont {Celine}\ \bibnamefont
  {Degrande}}, \bibinfo {author} {\bibfnamefont {Fabio}\ \bibnamefont
  {Maltoni}}, \bibinfo {author} {\bibfnamefont {Jian}\ \bibnamefont {Wang}}, \
  and\ \bibinfo {author} {\bibfnamefont {Cen}\ \bibnamefont {Zhang}},\
  }\bibfield  {title} {\enquote {\bibinfo {title} {{Automatic computations at
  next-to-leading order in QCD for top-quark flavor-changing neutral
  processes}},}\ }\href {\doibase 10.1103/PhysRevD.91.034024} {\bibfield
  {journal} {\bibinfo  {journal} {Phys. Rev.}\ }\textbf {\bibinfo {volume}
  {D91}},\ \bibinfo {pages} {034024} (\bibinfo {year} {2015})},\ \Eprint
  {http://arxiv.org/abs/1412.5594}{arXiv:1412.5594 [hep-ph]}\BibitemShut
  {NoStop}%
\bibitem [{\citenamefont {Durieux}\ \emph {et~al.}(2015)\citenamefont
  {Durieux}, \citenamefont {Maltoni},\ and\ \citenamefont
  {Zhang}}]{Durieux:2014xla}%
  \BibitemOpen
  \bibfield  {author} {\bibinfo {author} {\bibfnamefont {Gauthier}\
  \bibnamefont {Durieux}}, \bibinfo {author} {\bibfnamefont {Fabio}\
  \bibnamefont {Maltoni}}, \ and\ \bibinfo {author} {\bibfnamefont {Cen}\
  \bibnamefont {Zhang}},\ }\bibfield  {title} {\enquote {\bibinfo {title}
  {{Global approach to top-quark flavor-changing interactions}},}\ }\href
  {\doibase 10.1103/PhysRevD.91.074017} {\bibfield  {journal} {\bibinfo
  {journal} {Phys. Rev.}\ }\textbf {\bibinfo {volume} {D91}},\ \bibinfo {pages}
  {074017} (\bibinfo {year} {2015})},\ \Eprint
  {http://arxiv.org/abs/1412.7166}{arXiv:1412.7166 [hep-ph]}\BibitemShut
  {NoStop}%
\bibitem [{Top()}]{TopFCNC}%
  \BibitemOpen
  \href@noop {} {}\bibinfo {note}
  {\url{https://feynrules.irmp.ucl.ac.be/wiki/TopFCNC}}\BibitemShut {NoStop}%
\bibitem [{\citenamefont {Buarque~Franzosi}\ and\ \citenamefont
  {Zhang}(2015)}]{Franzosi:2015osa}%
  \BibitemOpen
  \bibfield  {author} {\bibinfo {author} {\bibfnamefont {Diogo}\ \bibnamefont
  {Buarque~Franzosi}}\ and\ \bibinfo {author} {\bibfnamefont {Cen}\
  \bibnamefont {Zhang}},\ }\bibfield  {title} {\enquote {\bibinfo {title}
  {{Probing the top-quark chromomagnetic dipole moment at next-to-leading order
  in QCD}},}\ }\href {\doibase 10.1103/PhysRevD.91.114010} {\bibfield
  {journal} {\bibinfo  {journal} {Phys. Rev.}\ }\textbf {\bibinfo {volume}
  {D91}},\ \bibinfo {pages} {114010} (\bibinfo {year} {2015})},\ \Eprint
  {http://arxiv.org/abs/1503.08841}{arXiv:1503.08841 [hep-ph]}\BibitemShut
  {NoStop}%
\bibitem [{\citenamefont {Maltoni}\ \emph {et~al.}(2016)\citenamefont
  {Maltoni}, \citenamefont {Vryonidou},\ and\ \citenamefont
  {Zhang}}]{Maltoni:2016yxb}%
  \BibitemOpen
  \bibfield  {author} {\bibinfo {author} {\bibfnamefont {Fabio}\ \bibnamefont
  {Maltoni}}, \bibinfo {author} {\bibfnamefont {Eleni}\ \bibnamefont
  {Vryonidou}}, \ and\ \bibinfo {author} {\bibfnamefont {Cen}\ \bibnamefont
  {Zhang}},\ }\bibfield  {title} {\enquote {\bibinfo {title} {{Higgs production
  in association with a top-antitop pair in the Standard Model Effective Field
  Theory at NLO in QCD}},}\ }\href {\doibase 10.1007/JHEP10(2016)123}
  {\bibfield  {journal} {\bibinfo  {journal} {JHEP}\ }\textbf {\bibinfo
  {volume} {10}},\ \bibinfo {pages} {123} (\bibinfo {year} {2016})},\ \Eprint
  {http://arxiv.org/abs/1607.05330}{arXiv:1607.05330 [hep-ph]}\BibitemShut
  {NoStop}%
\bibitem [{\citenamefont {Bessidskaia~Bylund}\ \emph
  {et~al.}(2016)\citenamefont {Bessidskaia~Bylund}, \citenamefont {Maltoni},
  \citenamefont {Tsinikos}, \citenamefont {Vryonidou},\ and\ \citenamefont
  {Zhang}}]{Bylund:2016phk}%
  \BibitemOpen
  \bibfield  {author} {\bibinfo {author} {\bibfnamefont {Olga}\ \bibnamefont
  {Bessidskaia~Bylund}}, \bibinfo {author} {\bibfnamefont {Fabio}\ \bibnamefont
  {Maltoni}}, \bibinfo {author} {\bibfnamefont {Ioannis}\ \bibnamefont
  {Tsinikos}}, \bibinfo {author} {\bibfnamefont {Eleni}\ \bibnamefont
  {Vryonidou}}, \ and\ \bibinfo {author} {\bibfnamefont {Cen}\ \bibnamefont
  {Zhang}},\ }\bibfield  {title} {\enquote {\bibinfo {title} {{Probing top
  quark neutral couplings in the Standard Model Effective Field Theory at NLO
  in QCD}},}\ }\href {\doibase 10.1007/JHEP05(2016)052} {\bibfield  {journal}
  {\bibinfo  {journal} {JHEP}\ }\textbf {\bibinfo {volume} {05}},\ \bibinfo
  {pages} {052} (\bibinfo {year} {2016})},\ \Eprint
  {http://arxiv.org/abs/1601.08193}{arXiv:1601.08193 [hep-ph]}\BibitemShut
  {NoStop}%
\bibitem [{\citenamefont {Zhang}(2016)}]{Zhang:2016omx}%
  \BibitemOpen
  \bibfield  {author} {\bibinfo {author} {\bibfnamefont {Cen}\ \bibnamefont
  {Zhang}},\ }\bibfield  {title} {\enquote {\bibinfo {title} {{Single Top
  Production at Next-to-Leading Order in the Standard Model Effective Field
  Theory}},}\ }\href {\doibase 10.1103/PhysRevLett.116.162002} {\bibfield
  {journal} {\bibinfo  {journal} {Phys. Rev. Lett.}\ }\textbf {\bibinfo
  {volume} {116}},\ \bibinfo {pages} {162002} (\bibinfo {year} {2016})},\
  \Eprint {http://arxiv.org/abs/1601.06163}{arXiv:1601.06163
  [hep-ph]}\BibitemShut {NoStop}%
\bibitem [{\citenamefont {de~Beurs}\ \emph {et~al.}(2018)\citenamefont
  {de~Beurs}, \citenamefont {Laenen}, \citenamefont {Vreeswijk},\ and\
  \citenamefont {Vryonidou}}]{deBeurs:2018pvs}%
  \BibitemOpen
  \bibfield  {author} {\bibinfo {author} {\bibfnamefont {M.}~\bibnamefont
  {de~Beurs}}, \bibinfo {author} {\bibfnamefont {E.}~\bibnamefont {Laenen}},
  \bibinfo {author} {\bibfnamefont {M.}~\bibnamefont {Vreeswijk}}, \ and\
  \bibinfo {author} {\bibfnamefont {E.}~\bibnamefont {Vryonidou}},\ }\bibfield
  {title} {\enquote {\bibinfo {title} {{Effective operators in $t$-channel
  single top production and decay}},}\ }\href {\doibase
  10.1140/epjc/s10052-018-6399-3} {\bibfield  {journal} {\bibinfo  {journal}
  {Eur.\ Phys.\ J.\ C}\ }\textbf {\bibinfo {volume} {78}},\ \bibinfo {pages}
  {919} (\bibinfo {year} {2018})},\ \Eprint
  {http://arxiv.org/abs/1807.03576}{arXiv:1807.03576 [hep-ph]}\BibitemShut
  {NoStop}%
\bibitem [{\citenamefont {Deutschmann}\ \emph {et~al.}(2017)\citenamefont
  {Deutschmann}, \citenamefont {Duhr}, \citenamefont {Maltoni},\ and\
  \citenamefont {Vryonidou}}]{Deutschmann:2017qum}%
  \BibitemOpen
  \bibfield  {author} {\bibinfo {author} {\bibfnamefont {Nicolas}\ \bibnamefont
  {Deutschmann}}, \bibinfo {author} {\bibfnamefont {Claude}\ \bibnamefont
  {Duhr}}, \bibinfo {author} {\bibfnamefont {Fabio}\ \bibnamefont {Maltoni}}, \
  and\ \bibinfo {author} {\bibfnamefont {Eleni}\ \bibnamefont {Vryonidou}},\
  }\bibfield  {title} {\enquote {\bibinfo {title} {{Gluon-fusion Higgs
  production in the Standard Model Effective Field Theory}},}\ }\href {\doibase
  10.1007/JHEP12(2017)063} {\bibfield  {journal} {\bibinfo  {journal} {JHEP}\
  }\textbf {\bibinfo {volume} {12}},\ \bibinfo {pages} {063} (\bibinfo {year}
  {2017})},\ \bibinfo {note} {[Erratum: JHEP 02, 159 (2018)]},\ \Eprint
  {http://arxiv.org/abs/1708.00460}{arXiv:1708.00460 [hep-ph]}\BibitemShut
  {NoStop}%
\bibitem [{\citenamefont {Hirschi}\ \emph {et~al.}(2018)\citenamefont
  {Hirschi}, \citenamefont {Maltoni}, \citenamefont {Tsinikos},\ and\
  \citenamefont {Vryonidou}}]{Hirschi:2018etq}%
  \BibitemOpen
  \bibfield  {author} {\bibinfo {author} {\bibfnamefont {Valentin}\
  \bibnamefont {Hirschi}}, \bibinfo {author} {\bibfnamefont {Fabio}\
  \bibnamefont {Maltoni}}, \bibinfo {author} {\bibfnamefont {Ioannis}\
  \bibnamefont {Tsinikos}}, \ and\ \bibinfo {author} {\bibfnamefont {Eleni}\
  \bibnamefont {Vryonidou}},\ }\bibfield  {title} {\enquote {\bibinfo {title}
  {{Constraining anomalous gluon self-interactions at the LHC: a
  reappraisal}},}\ }\href {\doibase 10.1007/JHEP07(2018)093} {\bibfield
  {journal} {\bibinfo  {journal} {JHEP}\ }\textbf {\bibinfo {volume} {07}},\
  \bibinfo {pages} {093} (\bibinfo {year} {2018})},\ \Eprint
  {http://arxiv.org/abs/1806.04696}{arXiv:1806.04696 [hep-ph]}\BibitemShut
  {NoStop}%
\bibitem [{\citenamefont {Degrande}\ \emph {et~al.}(2017)\citenamefont
  {Degrande}, \citenamefont {Fuks}, \citenamefont {Mawatari}, \citenamefont
  {Mimasu},\ and\ \citenamefont {Sanz}}]{Degrande:2016dqg}%
  \BibitemOpen
  \bibfield  {author} {\bibinfo {author} {\bibfnamefont {C\'eline}\
  \bibnamefont {Degrande}}, \bibinfo {author} {\bibfnamefont {Benjamin}\
  \bibnamefont {Fuks}}, \bibinfo {author} {\bibfnamefont {Kentarou}\
  \bibnamefont {Mawatari}}, \bibinfo {author} {\bibfnamefont {Ken}\
  \bibnamefont {Mimasu}}, \ and\ \bibinfo {author} {\bibfnamefont {Veronica}\
  \bibnamefont {Sanz}},\ }\bibfield  {title} {\enquote {\bibinfo {title}
  {{Electroweak Higgs boson production in the standard model effective field
  theory beyond leading order in QCD}},}\ }\href {\doibase
  10.1140/epjc/s10052-017-4793-x} {\bibfield  {journal} {\bibinfo  {journal}
  {Eur. Phys. J.}\ }\textbf {\bibinfo {volume} {C77}},\ \bibinfo {pages} {262}
  (\bibinfo {year} {2017})},\ \Eprint
  {http://arxiv.org/abs/1609.04833}{arXiv:1609.04833 [hep-ph]}\BibitemShut
  {NoStop}%
\bibitem [{\citenamefont {Degrande}\ \emph {et~al.}(2018)\citenamefont
  {Degrande}, \citenamefont {Maltoni}, \citenamefont {Mimasu}, \citenamefont
  {Vryonidou},\ and\ \citenamefont {Zhang}}]{Degrande:2018fog}%
  \BibitemOpen
  \bibfield  {author} {\bibinfo {author} {\bibfnamefont {C\'eline}\
  \bibnamefont {Degrande}}, \bibinfo {author} {\bibfnamefont {Fabio}\
  \bibnamefont {Maltoni}}, \bibinfo {author} {\bibfnamefont {Ken}\ \bibnamefont
  {Mimasu}}, \bibinfo {author} {\bibfnamefont {Eleni}\ \bibnamefont
  {Vryonidou}}, \ and\ \bibinfo {author} {\bibfnamefont {Cen}\ \bibnamefont
  {Zhang}},\ }\bibfield  {title} {\enquote {\bibinfo {title} {{Single-top
  associated production with a $Z$ or $H$ boson at the LHC: the SMEFT
  interpretation}},}\ }\href {\doibase 10.1007/JHEP10(2018)005} {\bibfield
  {journal} {\bibinfo  {journal} {JHEP}\ }\textbf {\bibinfo {volume} {10}},\
  \bibinfo {pages} {005} (\bibinfo {year} {2018})},\ \Eprint
  {http://arxiv.org/abs/1804.07773}{arXiv:1804.07773 [hep-ph]}\BibitemShut
  {NoStop}%
\bibitem [{\citenamefont {Brivio}\ \emph {et~al.}(2020)\citenamefont {Brivio},
  \citenamefont {Bruggisser}, \citenamefont {Maltoni}, \citenamefont
  {Moutafis}, \citenamefont {Plehn}, \citenamefont {Vryonidou}, \citenamefont
  {Westhoff},\ and\ \citenamefont {Zhang}}]{Brivio:2019ius}%
  \BibitemOpen
  \bibfield  {author} {\bibinfo {author} {\bibfnamefont {Ilaria}\ \bibnamefont
  {Brivio}}, \bibinfo {author} {\bibfnamefont {Sebastian}\ \bibnamefont
  {Bruggisser}}, \bibinfo {author} {\bibfnamefont {Fabio}\ \bibnamefont
  {Maltoni}}, \bibinfo {author} {\bibfnamefont {Rhea}\ \bibnamefont
  {Moutafis}}, \bibinfo {author} {\bibfnamefont {Tilman}\ \bibnamefont
  {Plehn}}, \bibinfo {author} {\bibfnamefont {Eleni}\ \bibnamefont
  {Vryonidou}}, \bibinfo {author} {\bibfnamefont {Susanne}\ \bibnamefont
  {Westhoff}}, \ and\ \bibinfo {author} {\bibfnamefont {C.}~\bibnamefont
  {Zhang}},\ }\bibfield  {title} {\enquote {\bibinfo {title} {{O new physics,
  where art thou? A global search in the top sector}},}\ }\href {\doibase
  10.1007/JHEP02(2020)131} {\bibfield  {journal} {\bibinfo  {journal} {JHEP}\
  }\textbf {\bibinfo {volume} {02}},\ \bibinfo {pages} {131} (\bibinfo {year}
  {2020})},\ \Eprint {http://arxiv.org/abs/1910.03606}{arXiv:1910.03606
  [hep-ph]}\BibitemShut {NoStop}%
\bibitem [{\citenamefont {Hartland}\ \emph {et~al.}(2019)\citenamefont
  {Hartland}, \citenamefont {Maltoni}, \citenamefont {Nocera}, \citenamefont
  {Rojo}, \citenamefont {Slade}, \citenamefont {Vryonidou},\ and\ \citenamefont
  {Zhang}}]{Hartland:2019bjb}%
  \BibitemOpen
  \bibfield  {author} {\bibinfo {author} {\bibfnamefont {Nathan~P.}\
  \bibnamefont {Hartland}}, \bibinfo {author} {\bibfnamefont {Fabio}\
  \bibnamefont {Maltoni}}, \bibinfo {author} {\bibfnamefont {Emanuele~R.}\
  \bibnamefont {Nocera}}, \bibinfo {author} {\bibfnamefont {Juan}\ \bibnamefont
  {Rojo}}, \bibinfo {author} {\bibfnamefont {Emma}\ \bibnamefont {Slade}},
  \bibinfo {author} {\bibfnamefont {Eleni}\ \bibnamefont {Vryonidou}}, \ and\
  \bibinfo {author} {\bibfnamefont {Cen}\ \bibnamefont {Zhang}},\ }\bibfield
  {title} {\enquote {\bibinfo {title} {{A Monte Carlo global analysis of the
  Standard Model Effective Field Theory: the top quark sector}},}\ }\href
  {\doibase 10.1007/JHEP04(2019)100} {\bibfield  {journal} {\bibinfo  {journal}
  {JHEP}\ }\textbf {\bibinfo {volume} {04}},\ \bibinfo {pages} {100} (\bibinfo
  {year} {2019})},\ \Eprint {http://arxiv.org/abs/1901.05965}{arXiv:1901.05965
  [hep-ph]}\BibitemShut {NoStop}%
\bibitem [{\citenamefont {Ball}\ \emph {et~al.}(2015)\citenamefont {Ball} \emph
  {et~al.}}]{Ball:2014uwa}%
  \BibitemOpen
  \bibfield  {author} {\bibinfo {author} {\bibfnamefont {Richard~D.}\
  \bibnamefont {Ball}} \emph {et~al.} (\bibinfo {collaboration} {NNPDF}),\
  }\bibfield  {title} {\enquote {\bibinfo {title} {{Parton distributions for
  the LHC Run II}},}\ }\href {\doibase 10.1007/JHEP04(2015)040} {\bibfield
  {journal} {\bibinfo  {journal} {JHEP}\ }\textbf {\bibinfo {volume} {04}},\
  \bibinfo {pages} {040} (\bibinfo {year} {2015})},\ \Eprint
  {http://arxiv.org/abs/1410.8849}{arXiv:1410.8849 [hep-ph]}\BibitemShut
  {NoStop}%
\bibitem [{\citenamefont {Sirunyan}\ \emph
  {et~al.}(2020{\natexlab{a}})\citenamefont {Sirunyan} \emph
  {et~al.}}]{Sirunyan:2019wxt}%
  \BibitemOpen
  \bibfield  {author} {\bibinfo {author} {\bibfnamefont {Albert~M}\
  \bibnamefont {Sirunyan}} \emph {et~al.} (\bibinfo {collaboration} {CMS}),\
  }\bibfield  {title} {\enquote {\bibinfo {title} {{Search for production of
  four top quarks in final states with same-sign or multiple leptons in
  proton-proton collisions at $\sqrt{s}=$ 13 TeV}},}\ }\href {\doibase
  10.1140/epjc/s10052-019-7593-7} {\bibfield  {journal} {\bibinfo  {journal}
  {Eur. Phys. J. C}\ }\textbf {\bibinfo {volume} {80}},\ \bibinfo {pages} {75}
  (\bibinfo {year} {2020}{\natexlab{a}})},\ \Eprint
  {http://arxiv.org/abs/1908.06463}{arXiv:1908.06463 [hep-ex]}\BibitemShut
  {NoStop}%
\bibitem [{\citenamefont {Aad}\ \emph {et~al.}(2020)\citenamefont {Aad} \emph
  {et~al.}}]{Aad:2020klt}%
  \BibitemOpen
  \bibfield  {author} {\bibinfo {author} {\bibfnamefont {Georges}\ \bibnamefont
  {Aad}} \emph {et~al.} (\bibinfo {collaboration} {ATLAS}),\ }\bibfield
  {title} {\enquote {\bibinfo {title} {{Evidence for $t\bar{t}t\bar{t}$
  production in the multilepton final state in proton-proton collisions at
  $\sqrt{s}$=13 TeV with the ATLAS detector}},}\ }\href@noop {} {\bibfield
  {journal} {\bibinfo  {journal} {Eur. Phys. J.}\ }\textbf {\bibinfo {volume}
  {C80}},\ \bibinfo {pages} {1085} (\bibinfo {year} {2020})},\ \Eprint
  {http://arxiv.org/abs/2007.14858}{arXiv:2007.14858 [hep-ex]}\BibitemShut
  {NoStop}%
\bibitem [{\citenamefont {Sirunyan}\ \emph
  {et~al.}(2020{\natexlab{b}})\citenamefont {Sirunyan} \emph
  {et~al.}}]{Sirunyan:2020cjp}%
  \BibitemOpen
  \bibfield  {author} {\bibinfo {author} {\bibfnamefont {Albert~M}\
  \bibnamefont {Sirunyan}} \emph {et~al.} (\bibinfo {collaboration} {CMS}),\
  }\bibfield  {title} {\enquote {\bibinfo {title} {{Observation of the
  Production of Three Massive Gauge Bosons at $\sqrt {s}$ =13\,\,TeV}},}\
  }\href {\doibase 10.1103/PhysRevLett.125.151802} {\bibfield  {journal}
  {\bibinfo  {journal} {Phys. Rev. Lett.}\ }\textbf {\bibinfo {volume} {125}},\
  \bibinfo {pages} {151802} (\bibinfo {year} {2020}{\natexlab{b}})},\ \Eprint
  {http://arxiv.org/abs/2006.11191}{arXiv:2006.11191 [hep-ex]}\BibitemShut
  {NoStop}%
\bibitem [{\citenamefont {Baglio}\ \emph {et~al.}(2017)\citenamefont {Baglio},
  \citenamefont {Dawson},\ and\ \citenamefont {Lewis}}]{Baglio:2017bfe}%
  \BibitemOpen
  \bibfield  {author} {\bibinfo {author} {\bibfnamefont {Julien}\ \bibnamefont
  {Baglio}}, \bibinfo {author} {\bibfnamefont {Sally}\ \bibnamefont {Dawson}},
  \ and\ \bibinfo {author} {\bibfnamefont {Ian~M.}\ \bibnamefont {Lewis}},\
  }\bibfield  {title} {\enquote {\bibinfo {title} {{An NLO QCD effective field
  theory analysis of $W^+W^-$ production at the LHC including fermionic
  operators}},}\ }\href {\doibase 10.1103/PhysRevD.96.073003} {\bibfield
  {journal} {\bibinfo  {journal} {Phys. Rev. D}\ }\textbf {\bibinfo {volume}
  {96}},\ \bibinfo {pages} {073003} (\bibinfo {year} {2017})},\ \Eprint
  {http://arxiv.org/abs/1708.03332}{arXiv:1708.03332 [hep-ph]}\BibitemShut
  {NoStop}%
\bibitem [{\citenamefont {Baglio}\ \emph {et~al.}(2019)\citenamefont {Baglio},
  \citenamefont {Dawson},\ and\ \citenamefont {Homiller}}]{Baglio:2019uty}%
  \BibitemOpen
  \bibfield  {author} {\bibinfo {author} {\bibfnamefont {Julien}\ \bibnamefont
  {Baglio}}, \bibinfo {author} {\bibfnamefont {Sally}\ \bibnamefont {Dawson}},
  \ and\ \bibinfo {author} {\bibfnamefont {Samuel}\ \bibnamefont {Homiller}},\
  }\bibfield  {title} {\enquote {\bibinfo {title} {{QCD corrections in Standard
  Model EFT fits to $WZ$ and $WW$ production}},}\ }\href {\doibase
  10.1103/PhysRevD.100.113010} {\bibfield  {journal} {\bibinfo  {journal}
  {Phys. Rev. D}\ }\textbf {\bibinfo {volume} {100}},\ \bibinfo {pages}
  {113010} (\bibinfo {year} {2019})},\ \Eprint
  {http://arxiv.org/abs/1909.11576}{arXiv:1909.11576 [hep-ph]}\BibitemShut
  {NoStop}%
\bibitem [{\citenamefont {Azatov}\ \emph {et~al.}(2015)\citenamefont {Azatov},
  \citenamefont {Grojean}, \citenamefont {Paul},\ and\ \citenamefont
  {Salvioni}}]{Azatov:2014jga}%
  \BibitemOpen
  \bibfield  {author} {\bibinfo {author} {\bibfnamefont {Aleksandr}\
  \bibnamefont {Azatov}}, \bibinfo {author} {\bibfnamefont {Christophe}\
  \bibnamefont {Grojean}}, \bibinfo {author} {\bibfnamefont {Ayan}\
  \bibnamefont {Paul}}, \ and\ \bibinfo {author} {\bibfnamefont {Ennio}\
  \bibnamefont {Salvioni}},\ }\bibfield  {title} {\enquote {\bibinfo {title}
  {{Taming the off-shell Higgs boson}},}\ }\href {\doibase
  10.1134/S1063776115030140} {\bibfield  {journal} {\bibinfo  {journal} {Zh.
  Eksp. Teor. Fiz.}\ }\textbf {\bibinfo {volume} {147}},\ \bibinfo {pages}
  {410--425} (\bibinfo {year} {2015})},\ \Eprint
  {http://arxiv.org/abs/1406.6338}{arXiv:1406.6338 [hep-ph]}\BibitemShut
  {NoStop}%
\bibitem [{\citenamefont {Hirschi}\ and\ \citenamefont
  {Mattelaer}(2015)}]{Hirschi:2015iia}%
  \BibitemOpen
  \bibfield  {author} {\bibinfo {author} {\bibfnamefont {Valentin}\
  \bibnamefont {Hirschi}}\ and\ \bibinfo {author} {\bibfnamefont {Olivier}\
  \bibnamefont {Mattelaer}},\ }\bibfield  {title} {\enquote {\bibinfo {title}
  {{Automated event generation for loop-induced processes}},}\ }\href {\doibase
  10.1007/JHEP10(2015)146} {\bibfield  {journal} {\bibinfo  {journal} {JHEP}\
  }\textbf {\bibinfo {volume} {10}},\ \bibinfo {pages} {146} (\bibinfo {year}
  {2015})},\ \Eprint {http://arxiv.org/abs/1507.00020}{arXiv:1507.00020
  [hep-ph]}\BibitemShut {NoStop}%
\bibitem [{\citenamefont {Azatov}\ \emph
  {et~al.}(2017{\natexlab{a}})\citenamefont {Azatov}, \citenamefont {Contino},
  \citenamefont {Machado},\ and\ \citenamefont {Riva}}]{Azatov:2016sqh}%
  \BibitemOpen
  \bibfield  {author} {\bibinfo {author} {\bibfnamefont {Aleksandr}\
  \bibnamefont {Azatov}}, \bibinfo {author} {\bibfnamefont {Roberto}\
  \bibnamefont {Contino}}, \bibinfo {author} {\bibfnamefont {Camila~S.}\
  \bibnamefont {Machado}}, \ and\ \bibinfo {author} {\bibfnamefont {Francesco}\
  \bibnamefont {Riva}},\ }\bibfield  {title} {\enquote {\bibinfo {title}
  {{Helicity selection rules and noninterference for BSM amplitudes}},}\ }\href
  {\doibase 10.1103/PhysRevD.95.065014} {\bibfield  {journal} {\bibinfo
  {journal} {Phys. Rev. D}\ }\textbf {\bibinfo {volume} {95}},\ \bibinfo
  {pages} {065014} (\bibinfo {year} {2017}{\natexlab{a}})},\ \Eprint
  {http://arxiv.org/abs/1607.05236}{arXiv:1607.05236 [hep-ph]}\BibitemShut
  {NoStop}%
\bibitem [{\citenamefont {Azatov}\ \emph
  {et~al.}(2017{\natexlab{b}})\citenamefont {Azatov}, \citenamefont
  {Elias-Miro}, \citenamefont {Reyimuaji},\ and\ \citenamefont
  {Venturini}}]{Azatov:2017kzw}%
  \BibitemOpen
  \bibfield  {author} {\bibinfo {author} {\bibfnamefont {A.}~\bibnamefont
  {Azatov}}, \bibinfo {author} {\bibfnamefont {J.}~\bibnamefont {Elias-Miro}},
  \bibinfo {author} {\bibfnamefont {Y.}~\bibnamefont {Reyimuaji}}, \ and\
  \bibinfo {author} {\bibfnamefont {E.}~\bibnamefont {Venturini}},\ }\bibfield
  {title} {\enquote {\bibinfo {title} {{Novel measurements of anomalous triple
  gauge couplings for the LHC}},}\ }\href {\doibase 10.1007/JHEP10(2017)027}
  {\bibfield  {journal} {\bibinfo  {journal} {JHEP}\ }\textbf {\bibinfo
  {volume} {10}},\ \bibinfo {pages} {027} (\bibinfo {year}
  {2017}{\natexlab{b}})},\ \Eprint
  {http://arxiv.org/abs/1707.08060}{arXiv:1707.08060 [hep-ph]}\BibitemShut
  {NoStop}%
\bibitem [{\citenamefont {Panico}\ \emph {et~al.}(2018)\citenamefont {Panico},
  \citenamefont {Riva},\ and\ \citenamefont {Wulzer}}]{Panico:2017frx}%
  \BibitemOpen
  \bibfield  {author} {\bibinfo {author} {\bibfnamefont {Giuliano}\
  \bibnamefont {Panico}}, \bibinfo {author} {\bibfnamefont {Francesco}\
  \bibnamefont {Riva}}, \ and\ \bibinfo {author} {\bibfnamefont {Andrea}\
  \bibnamefont {Wulzer}},\ }\bibfield  {title} {\enquote {\bibinfo {title}
  {{Diboson Interference Resurrection}},}\ }\href {\doibase
  10.1016/j.physletb.2017.11.068} {\bibfield  {journal} {\bibinfo  {journal}
  {Phys. Lett. B}\ }\textbf {\bibinfo {volume} {776}},\ \bibinfo {pages}
  {473--480} (\bibinfo {year} {2018})},\ \Eprint
  {http://arxiv.org/abs/1708.07823}{arXiv:1708.07823 [hep-ph]}\BibitemShut
  {NoStop}%
\bibitem [{\citenamefont {Degrassi}\ \emph {et~al.}(2016)\citenamefont
  {Degrassi}, \citenamefont {Giardino}, \citenamefont {Maltoni},\ and\
  \citenamefont {Pagani}}]{Degrassi:2016wml}%
  \BibitemOpen
  \bibfield  {author} {\bibinfo {author} {\bibfnamefont {Giuseppe}\
  \bibnamefont {Degrassi}}, \bibinfo {author} {\bibfnamefont {Pier~Paolo}\
  \bibnamefont {Giardino}}, \bibinfo {author} {\bibfnamefont {Fabio}\
  \bibnamefont {Maltoni}}, \ and\ \bibinfo {author} {\bibfnamefont {Davide}\
  \bibnamefont {Pagani}},\ }\bibfield  {title} {\enquote {\bibinfo {title}
  {{Probing the Higgs self coupling via single Higgs production at the LHC}},}\
  }\href {\doibase 10.1007/JHEP12(2016)080} {\bibfield  {journal} {\bibinfo
  {journal} {JHEP}\ }\textbf {\bibinfo {volume} {12}},\ \bibinfo {pages} {080}
  (\bibinfo {year} {2016})},\ \Eprint
  {http://arxiv.org/abs/1607.04251}{arXiv:1607.04251 [hep-ph]}\BibitemShut
  {NoStop}%
\bibitem [{\citenamefont {Abada}\ \emph {et~al.}(2019)\citenamefont {Abada}
  \emph {et~al.}}]{Abada:2019lih}%
  \BibitemOpen
  \bibfield  {author} {\bibinfo {author} {\bibfnamefont {A.}~\bibnamefont
  {Abada}} \emph {et~al.} (\bibinfo {collaboration} {FCC}),\ }\bibfield
  {title} {\enquote {\bibinfo {title} {{FCC Physics Opportunities}: {Future
  Circular Collider Conceptual Design Report Volume 1}},}\ }\href {\doibase
  10.1140/epjc/s10052-019-6904-3} {\bibfield  {journal} {\bibinfo  {journal}
  {Eur. Phys. J. C}\ }\textbf {\bibinfo {volume} {79}},\ \bibinfo {pages} {474}
  (\bibinfo {year} {2019})}\BibitemShut {NoStop}%
\bibitem [{\citenamefont {Mangano}\ \emph {et~al.}(2020)\citenamefont
  {Mangano}, \citenamefont {Ortona},\ and\ \citenamefont
  {Selvaggi}}]{Mangano:2020sao}%
  \BibitemOpen
  \bibfield  {author} {\bibinfo {author} {\bibfnamefont {Michelangelo~L.}\
  \bibnamefont {Mangano}}, \bibinfo {author} {\bibfnamefont {Giacomo}\
  \bibnamefont {Ortona}}, \ and\ \bibinfo {author} {\bibfnamefont {Michele}\
  \bibnamefont {Selvaggi}},\ }\bibfield  {title} {\enquote {\bibinfo {title}
  {{Measuring the Higgs self-coupling via Higgs-pair production at a 100 TeV
  p-p collider}},}\ }\href {\doibase 10.1140/epjc/s10052-020-08595-3}
  {\bibfield  {journal} {\bibinfo  {journal} {Eur. Phys. J.}\ }\textbf
  {\bibinfo {volume} {C80}},\ \bibinfo {pages} {1030} (\bibinfo {year}
  {2020})},\ \Eprint {http://arxiv.org/abs/2004.03505}{arXiv:2004.03505
  [hep-ph]}\BibitemShut {NoStop}%
\bibitem [{\citenamefont {Papaefstathiou}\ \emph {et~al.}(2019)\citenamefont
  {Papaefstathiou}, \citenamefont {Tetlalmatzi-Xolocotzi},\ and\ \citenamefont
  {Zaro}}]{Papaefstathiou:2019ofh}%
  \BibitemOpen
  \bibfield  {author} {\bibinfo {author} {\bibfnamefont {Andreas}\ \bibnamefont
  {Papaefstathiou}}, \bibinfo {author} {\bibfnamefont {Gilberto}\ \bibnamefont
  {Tetlalmatzi-Xolocotzi}}, \ and\ \bibinfo {author} {\bibfnamefont {Marco}\
  \bibnamefont {Zaro}},\ }\bibfield  {title} {\enquote {\bibinfo {title}
  {{Triple Higgs boson production to six $b$-jets at a 100 TeV proton
  collider}},}\ }\href {\doibase 10.1140/epjc/s10052-019-7457-1} {\bibfield
  {journal} {\bibinfo  {journal} {Eur. Phys. J. C}\ }\textbf {\bibinfo {volume}
  {79}},\ \bibinfo {pages} {947} (\bibinfo {year} {2019})},\ \Eprint
  {http://arxiv.org/abs/1909.09166}{arXiv:1909.09166 [hep-ph]}\BibitemShut
  {NoStop}%
\bibitem [{\citenamefont {Frederix}\ \emph {et~al.}(2018)\citenamefont
  {Frederix}, \citenamefont {Frixione}, \citenamefont {Hirschi}, \citenamefont
  {Pagani}, \citenamefont {Shao},\ and\ \citenamefont
  {Zaro}}]{Frederix:2018nkq}%
  \BibitemOpen
  \bibfield  {author} {\bibinfo {author} {\bibfnamefont {R.}~\bibnamefont
  {Frederix}}, \bibinfo {author} {\bibfnamefont {S.}~\bibnamefont {Frixione}},
  \bibinfo {author} {\bibfnamefont {V.}~\bibnamefont {Hirschi}}, \bibinfo
  {author} {\bibfnamefont {D.}~\bibnamefont {Pagani}}, \bibinfo {author}
  {\bibfnamefont {H.-S.}\ \bibnamefont {Shao}}, \ and\ \bibinfo {author}
  {\bibfnamefont {M.}~\bibnamefont {Zaro}},\ }\bibfield  {title} {\enquote
  {\bibinfo {title} {{The automation of next-to-leading order electroweak
  calculations}},}\ }\href {\doibase 10.1007/JHEP07(2018)185} {\bibfield
  {journal} {\bibinfo  {journal} {JHEP}\ }\textbf {\bibinfo {volume} {07}},\
  \bibinfo {pages} {185} (\bibinfo {year} {2018})},\ \Eprint
  {http://arxiv.org/abs/1804.10017}{arXiv:1804.10017 [hep-ph]}\BibitemShut
  {NoStop}%
\bibitem [{\citenamefont {Hartmann}\ and\ \citenamefont
  {Trott}(2015)}]{Hartmann:2015oia}%
  \BibitemOpen
  \bibfield  {author} {\bibinfo {author} {\bibfnamefont {Christine}\
  \bibnamefont {Hartmann}}\ and\ \bibinfo {author} {\bibfnamefont {Michael}\
  \bibnamefont {Trott}},\ }\bibfield  {title} {\enquote {\bibinfo {title} {{On
  one-loop corrections in the standard model effective field theory; the
  $\Gamma(h \rightarrow \gamma \, \gamma)$ case}},}\ }\href {\doibase
  10.1007/JHEP07(2015)151} {\bibfield  {journal} {\bibinfo  {journal} {JHEP}\
  }\textbf {\bibinfo {volume} {07}},\ \bibinfo {pages} {151} (\bibinfo {year}
  {2015})},\ \Eprint {http://arxiv.org/abs/1505.02646}{arXiv:1505.02646
  [hep-ph]}\BibitemShut {NoStop}%
\bibitem [{\citenamefont {Ghezzi}\ \emph {et~al.}(2015)\citenamefont {Ghezzi},
  \citenamefont {Gomez-Ambrosio}, \citenamefont {Passarino},\ and\
  \citenamefont {Uccirati}}]{Ghezzi:2015vva}%
  \BibitemOpen
  \bibfield  {author} {\bibinfo {author} {\bibfnamefont {Margherita}\
  \bibnamefont {Ghezzi}}, \bibinfo {author} {\bibfnamefont {Raquel}\
  \bibnamefont {Gomez-Ambrosio}}, \bibinfo {author} {\bibfnamefont {Giampiero}\
  \bibnamefont {Passarino}}, \ and\ \bibinfo {author} {\bibfnamefont {Sandro}\
  \bibnamefont {Uccirati}},\ }\bibfield  {title} {\enquote {\bibinfo {title}
  {{NLO Higgs effective field theory and $\kappa$-framework}},}\ }\href
  {\doibase 10.1007/JHEP07(2015)175} {\bibfield  {journal} {\bibinfo  {journal}
  {JHEP}\ }\textbf {\bibinfo {volume} {07}},\ \bibinfo {pages} {175} (\bibinfo
  {year} {2015})},\ \Eprint {http://arxiv.org/abs/1505.03706}{arXiv:1505.03706
  [hep-ph]}\BibitemShut {NoStop}%
\bibitem [{\citenamefont {Hartmann}\ \emph {et~al.}(2017)\citenamefont
  {Hartmann}, \citenamefont {Shepherd},\ and\ \citenamefont
  {Trott}}]{Hartmann:2016pil}%
  \BibitemOpen
  \bibfield  {author} {\bibinfo {author} {\bibfnamefont {Christine}\
  \bibnamefont {Hartmann}}, \bibinfo {author} {\bibfnamefont {William}\
  \bibnamefont {Shepherd}}, \ and\ \bibinfo {author} {\bibfnamefont {Michael}\
  \bibnamefont {Trott}},\ }\bibfield  {title} {\enquote {\bibinfo {title} {{The
  $Z$ decay width in the SMEFT: $y_t$ and $\lambda$ corrections at one
  loop}},}\ }\href {\doibase 10.1007/JHEP03(2017)060} {\bibfield  {journal}
  {\bibinfo  {journal} {JHEP}\ }\textbf {\bibinfo {volume} {03}},\ \bibinfo
  {pages} {060} (\bibinfo {year} {2017})},\ \Eprint
  {http://arxiv.org/abs/1611.09879}{arXiv:1611.09879 [hep-ph]}\BibitemShut
  {NoStop}%
\bibitem [{\citenamefont {Dawson}\ and\ \citenamefont
  {Giardino}(2018{\natexlab{a}})}]{Dawson:2018pyl}%
  \BibitemOpen
  \bibfield  {author} {\bibinfo {author} {\bibfnamefont {Sally}\ \bibnamefont
  {Dawson}}\ and\ \bibinfo {author} {\bibfnamefont {Pier~Paolo}\ \bibnamefont
  {Giardino}},\ }\bibfield  {title} {\enquote {\bibinfo {title} {{Higgs decays
  to $ZZ$ and $Z\gamma$ in the standard model effective field theory: An NLO
  analysis}},}\ }\href {\doibase 10.1103/PhysRevD.97.093003} {\bibfield
  {journal} {\bibinfo  {journal} {Phys. Rev. D}\ }\textbf {\bibinfo {volume}
  {97}},\ \bibinfo {pages} {093003} (\bibinfo {year} {2018}{\natexlab{a}})},\
  \Eprint {http://arxiv.org/abs/1801.01136}{arXiv:1801.01136
  [hep-ph]}\BibitemShut {NoStop}%
\bibitem [{\citenamefont {Dedes}\ \emph {et~al.}(2018)\citenamefont {Dedes},
  \citenamefont {Paraskevas}, \citenamefont {Rosiek}, \citenamefont {Suxho},\
  and\ \citenamefont {Trifyllis}}]{Dedes:2018seb}%
  \BibitemOpen
  \bibfield  {author} {\bibinfo {author} {\bibfnamefont {A.}~\bibnamefont
  {Dedes}}, \bibinfo {author} {\bibfnamefont {M.}~\bibnamefont {Paraskevas}},
  \bibinfo {author} {\bibfnamefont {J.}~\bibnamefont {Rosiek}}, \bibinfo
  {author} {\bibfnamefont {K.}~\bibnamefont {Suxho}}, \ and\ \bibinfo {author}
  {\bibfnamefont {L.}~\bibnamefont {Trifyllis}},\ }\bibfield  {title} {\enquote
  {\bibinfo {title} {{The decay $h\to \gamma\gamma$ in the Standard-Model
  Effective Field Theory}},}\ }\href {\doibase 10.1007/JHEP08(2018)103}
  {\bibfield  {journal} {\bibinfo  {journal} {JHEP}\ }\textbf {\bibinfo
  {volume} {08}},\ \bibinfo {pages} {103} (\bibinfo {year} {2018})},\ \Eprint
  {http://arxiv.org/abs/1805.00302}{arXiv:1805.00302 [hep-ph]}\BibitemShut
  {NoStop}%
\bibitem [{\citenamefont {Dawson}\ and\ \citenamefont
  {Giardino}(2018{\natexlab{b}})}]{Dawson:2018liq}%
  \BibitemOpen
  \bibfield  {author} {\bibinfo {author} {\bibfnamefont {Sally}\ \bibnamefont
  {Dawson}}\ and\ \bibinfo {author} {\bibfnamefont {Pier~Paolo}\ \bibnamefont
  {Giardino}},\ }\bibfield  {title} {\enquote {\bibinfo {title} {{Electroweak
  corrections to Higgs boson decays to $\gamma\gamma$ and $W^+W^-$ in standard
  model EFT}},}\ }\href {\doibase 10.1103/PhysRevD.98.095005} {\bibfield
  {journal} {\bibinfo  {journal} {Phys. Rev. D}\ }\textbf {\bibinfo {volume}
  {98}},\ \bibinfo {pages} {095005} (\bibinfo {year} {2018}{\natexlab{b}})},\
  \Eprint {http://arxiv.org/abs/1807.11504}{arXiv:1807.11504
  [hep-ph]}\BibitemShut {NoStop}%
\bibitem [{\citenamefont {Dawson}\ and\ \citenamefont
  {Ismail}(2018)}]{Dawson:2018jlg}%
  \BibitemOpen
  \bibfield  {author} {\bibinfo {author} {\bibfnamefont {Sally}\ \bibnamefont
  {Dawson}}\ and\ \bibinfo {author} {\bibfnamefont {Ahmed}\ \bibnamefont
  {Ismail}},\ }\bibfield  {title} {\enquote {\bibinfo {title} {{Standard model
  EFT corrections to Z boson decays}},}\ }\href {\doibase
  10.1103/PhysRevD.98.093003} {\bibfield  {journal} {\bibinfo  {journal} {Phys.
  Rev. D}\ }\textbf {\bibinfo {volume} {98}},\ \bibinfo {pages} {093003}
  (\bibinfo {year} {2018})},\ \Eprint
  {http://arxiv.org/abs/1808.05948}{arXiv:1808.05948 [hep-ph]}\BibitemShut
  {NoStop}%
\bibitem [{\citenamefont {Dedes}\ \emph {et~al.}(2019)\citenamefont {Dedes},
  \citenamefont {Suxho},\ and\ \citenamefont {Trifyllis}}]{Dedes:2019bew}%
  \BibitemOpen
  \bibfield  {author} {\bibinfo {author} {\bibfnamefont {A.}~\bibnamefont
  {Dedes}}, \bibinfo {author} {\bibfnamefont {K.}~\bibnamefont {Suxho}}, \ and\
  \bibinfo {author} {\bibfnamefont {L.}~\bibnamefont {Trifyllis}},\ }\bibfield
  {title} {\enquote {\bibinfo {title} {{The decay $h\to Z \gamma$ in the
  Standard-Model Effective Field Theory}},}\ }\href {\doibase
  10.1007/JHEP06(2019)115} {\bibfield  {journal} {\bibinfo  {journal} {JHEP}\
  }\textbf {\bibinfo {volume} {06}},\ \bibinfo {pages} {115} (\bibinfo {year}
  {2019})},\ \Eprint {http://arxiv.org/abs/1903.12046}{arXiv:1903.12046
  [hep-ph]}\BibitemShut {NoStop}%
\bibitem [{\citenamefont {Cullen}\ \emph {et~al.}(2019)\citenamefont {Cullen},
  \citenamefont {Pecjak},\ and\ \citenamefont {Scott}}]{Cullen:2019nnr}%
  \BibitemOpen
  \bibfield  {author} {\bibinfo {author} {\bibfnamefont {Jonathan~M.}\
  \bibnamefont {Cullen}}, \bibinfo {author} {\bibfnamefont {Benjamin~D.}\
  \bibnamefont {Pecjak}}, \ and\ \bibinfo {author} {\bibfnamefont {Darren~J.}\
  \bibnamefont {Scott}},\ }\bibfield  {title} {\enquote {\bibinfo {title} {{NLO
  corrections to $h\to b\bar b$ decay in SMEFT}},}\ }\href {\doibase
  10.1007/JHEP08(2019)173} {\bibfield  {journal} {\bibinfo  {journal} {JHEP}\
  }\textbf {\bibinfo {volume} {08}},\ \bibinfo {pages} {173} (\bibinfo {year}
  {2019})},\ \Eprint {http://arxiv.org/abs/1904.06358}{arXiv:1904.06358
  [hep-ph]}\BibitemShut {NoStop}%
\bibitem [{\citenamefont {Dawson}\ and\ \citenamefont
  {Giardino}(2020)}]{Dawson:2019clf}%
  \BibitemOpen
  \bibfield  {author} {\bibinfo {author} {\bibfnamefont {Sally}\ \bibnamefont
  {Dawson}}\ and\ \bibinfo {author} {\bibfnamefont {Pier~Paolo}\ \bibnamefont
  {Giardino}},\ }\bibfield  {title} {\enquote {\bibinfo {title} {{Electroweak
  and QCD corrections to $Z$ and $W$ pole observables in the standard model
  EFT}},}\ }\href {\doibase 10.1103/PhysRevD.101.013001} {\bibfield  {journal}
  {\bibinfo  {journal} {Phys. Rev. D}\ }\textbf {\bibinfo {volume} {101}},\
  \bibinfo {pages} {013001} (\bibinfo {year} {2020})},\ \Eprint
  {http://arxiv.org/abs/1909.02000}{arXiv:1909.02000 [hep-ph]}\BibitemShut
  {NoStop}%
\bibitem [{\citenamefont {Martini}\ and\ \citenamefont
  {Schulze}(2020)}]{Martini:2019lsi}%
  \BibitemOpen
  \bibfield  {author} {\bibinfo {author} {\bibfnamefont {Till}\ \bibnamefont
  {Martini}}\ and\ \bibinfo {author} {\bibfnamefont {Markus}\ \bibnamefont
  {Schulze}},\ }\bibfield  {title} {\enquote {\bibinfo {title} {{Electroweak
  loops as a probe of new physics in $ t\overline{t} $ production at the
  LHC}},}\ }\href {\doibase 10.1007/JHEP04(2020)017} {\bibfield  {journal}
  {\bibinfo  {journal} {JHEP}\ }\textbf {\bibinfo {volume} {04}},\ \bibinfo
  {pages} {017} (\bibinfo {year} {2020})},\ \Eprint
  {http://arxiv.org/abs/1911.11244}{arXiv:1911.11244 [hep-ph]}\BibitemShut
  {NoStop}%
\bibitem [{\citenamefont {Cullen}\ and\ \citenamefont
  {Pecjak}(2020)}]{Cullen:2020zof}%
  \BibitemOpen
  \bibfield  {author} {\bibinfo {author} {\bibfnamefont {Jonathan~M.}\
  \bibnamefont {Cullen}}\ and\ \bibinfo {author} {\bibfnamefont {Benjamin~D.}\
  \bibnamefont {Pecjak}},\ }\bibfield  {title} {\enquote {\bibinfo {title}
  {{Higgs decay to fermion pairs at NLO in SMEFT}},}\ }\href {\doibase
  10.1007/JHEP11(2020)079} {\bibfield  {journal} {\bibinfo  {journal} {JHEP}\
  }\textbf {\bibinfo {volume} {11}},\ \bibinfo {pages} {079} (\bibinfo {year}
  {2020})},\ \Eprint {http://arxiv.org/abs/2007.15238}{arXiv:2007.15238
  [hep-ph]}\BibitemShut {NoStop}%
\bibitem [{\citenamefont {Vryonidou}\ and\ \citenamefont
  {Zhang}(2018)}]{Vryonidou:2018eyv}%
  \BibitemOpen
  \bibfield  {author} {\bibinfo {author} {\bibfnamefont {Eleni}\ \bibnamefont
  {Vryonidou}}\ and\ \bibinfo {author} {\bibfnamefont {Cen}\ \bibnamefont
  {Zhang}},\ }\bibfield  {title} {\enquote {\bibinfo {title} {{Dimension-six
  electroweak top-loop effects in Higgs production and decay}},}\ }\href
  {\doibase 10.1007/JHEP08(2018)036} {\bibfield  {journal} {\bibinfo  {journal}
  {JHEP}\ }\textbf {\bibinfo {volume} {08}},\ \bibinfo {pages} {036} (\bibinfo
  {year} {2018})},\ \Eprint {http://arxiv.org/abs/1804.09766}{arXiv:1804.09766
  [hep-ph]}\BibitemShut {NoStop}%
\bibitem [{\citenamefont {Durieux}\ \emph {et~al.}(2018)\citenamefont
  {Durieux}, \citenamefont {Gu}, \citenamefont {Vryonidou},\ and\ \citenamefont
  {Zhang}}]{Durieux:2018ggn}%
  \BibitemOpen
  \bibfield  {author} {\bibinfo {author} {\bibfnamefont {Gauthier}\
  \bibnamefont {Durieux}}, \bibinfo {author} {\bibfnamefont {Jiayin}\
  \bibnamefont {Gu}}, \bibinfo {author} {\bibfnamefont {Eleni}\ \bibnamefont
  {Vryonidou}}, \ and\ \bibinfo {author} {\bibfnamefont {Cen}\ \bibnamefont
  {Zhang}},\ }\bibfield  {title} {\enquote {\bibinfo {title} {{Probing
  top-quark couplings indirectly at Higgs factories}},}\ }\href {\doibase
  10.1088/1674-1137/42/12/123107} {\bibfield  {journal} {\bibinfo  {journal}
  {Chin.\ Phys.\ C}\ }\textbf {\bibinfo {volume} {42}},\ \bibinfo {pages}
  {123107} (\bibinfo {year} {2018})},\ \Eprint
  {http://arxiv.org/abs/1809.03520}{arXiv:1809.03520 [hep-ph]}\BibitemShut
  {NoStop}%
\end{thebibliography}%

\end{document}